%% file: main.tex
\documentclass[lettersize,journal]{IEEEtran}
\usepackage{amsmath,amsfonts}
\usepackage{algorithmic}
\usepackage{algorithm}
\usepackage{array}
\usepackage[caption=false,font=normalsize,labelfont=sf,textfont=sf]{subfig}
\usepackage{textcomp}
\usepackage{stfloats}
\usepackage{url}
\usepackage{verbatim}
\usepackage{graphicx}
\usepackage{cite}
\usepackage{amssymb}
\usepackage{url}
\usepackage{mathtools}
\usepackage{bm}
\usepackage{mathtools}
\usepackage{booktabs}
\usepackage{graphics}
\usepackage{wrapfig}
\usepackage[export]{adjustbox}
\usepackage{algorithm}
\usepackage{algorithmic}
\usepackage{multirow}
\usepackage{float}
\usepackage{soul}
\usepackage{hyperref}
\usepackage{booktabs}
\usepackage{tabularx}
\usepackage{makecell} 
\usepackage{balance}

\usepackage[numbers,sort&compress]{natbib}

\usepackage[dvipsnames, table, xcdraw]{xcolor} 
\usepackage{pifont} 
\newcommand{\cmark}{\ding{51}}%
\definecolor{bsRed}{rgb}{0.95, 0.0, 0.0}

\begin{document}

\title{Towards Generalized Source Tracing for Codec-Based Deepfake Speech}

\author{ Xuanjun Chen$^{1*}$, I-Ming Lin$^{2*}$ \thanks{*Equal Contribution; $\dagger$ Corresponding Author; Supported by the National Science and Technology Council, Taiwan (Grant NSTC 112-2634-F-002-005, Advanced Technologies for Designing Trustable AI Services).}, Lin Zhang$^{3}$, Haibin Wu$^{4\dagger}$, Hung-yi Lee$^{1}$, Jyh-Shing Roger Jang$^{2}$ \\
\textit{$^1$Graduate Institute of Communications Engineering, National Taiwan University} \\
\textit{$^2$Department of Computer Science and Information Engineering, National Taiwan University} \\
\textit{$^3$Center for Language and Speech Processing, Johns Hopkins University} \\
\textit{$^4$Independent Researcher}
}

\newcommand{\proofread}[1]{{\small\textcolor{orange}{\bf [#1]}}}
\newcommand{\units}[1]{{\small\textcolor{green}{\bf [#1]}}}
\newcommand{\zlin}[1]{{\textcolor{blue}{[#1]}}}

\maketitle
\input{sections/0.abstract}
\input{sections/1.intro}
\input{sections/2.background}

\input{sections/3.SASTNet}
\input{sections/4.setup}
\input{sections/5.main_results}
\input{sections/6.analysis_and_discussion}

\input{sections/7.conclusion}

\section{Acknowledgments}
We thank National Center for High-performance Computing (NCHC) of National Applied Research Laboratories (NARLabs) in Taiwan for providing computational and storage resources.

\section{Broader Impacts}
In recent years, general-purpose audio language models have become foundational for speech understanding \cite{huang2024dynamic, huang2025dynamicsuperb, yang2025towards}. However, even state‐of‐the‐art systems like GPT-4o still steer clear of deepfake detection because of safety restrictions \cite{lin2025preliminary}, underscoring their blind spots when it comes to synthetic audio threats. By contrast, SASTNet provides a targeted, interpretable framework specifically designed to trace codec‐based speech synthesis. We hope our approach sheds light on this critical problem and spurs further exploration in the field. 

\bibliographystyle{IEEEtran}
\bibliography{refs}

\end{document}

%% file: sections/0.abstract.tex
\begin{abstract}

Recent attempts at source tracing for codec-based deepfake speech (CodecFake), generated by neural audio codec–based speech generation (CoSG) models, have exhibited suboptimal performance. 
However, how to train source tracing models using simulated CoSG data while maintaining strong performance on real CoSG-generated audio remains an open challenge. In this paper, we show that models trained solely on codec-resynthesized data tend to overfit to non-speech regions and struggle to generalize to unseen content. 
To mitigate these challenges, we introduce the Semantic-Acoustic Source Tracing Network (SASTNet), which jointly leverages Whisper for semantic feature encoding and Wav2vec2 with AudioMAE for acoustic feature encoding. 
Our proposed SASTNet achieves state-of-the-art performance on the CoSG test set of \textit{CodecFake+} dataset, demonstrating its effectiveness for reliable source tracing. 

\end{abstract}
\begin{IEEEkeywords}
Anti-spoofing, source tracing, audio deepfake detection, neural audio codec, explainability
\end{IEEEkeywords}

%% file: sections/1.intro.tex
\section{Introduction}

Deepfake detection determines whether the given speech is a bona fide speech or a deepfake speech. 
Numerous challenges and datasets~\cite{li2024audio,wu2023defender} have been introduced, such as  ASVspoof \cite{wu15e_interspeech, kinnunen17_interspeech, todisco2019asvspoof, Liu_2023, Wang2024_ASVspoof5} and ADD \cite{yi2022add, yi2024add2023}. 
Recently, attention has shifted from merely detecting deepfake speech to tracing its source. Source tracing offers important insights into the origin or generation algorithm of a deepfake speech, which is especially valuable in out‐of‐domain detection scenarios. 
The simplest approach to source tracing \cite{borrelli2021synthetic} involves mapping deepfake utterances back to the specific attack model ID that generated them \cite{yi2024add2023}. 
To enhance generalization, the field has shifted toward attribute‐based source tracing \cite{Yan2022AnII, Zhang2023DistinguishingNS, zhu2022source, klein24_interspeech}. 
Rather than treating each deepfake system as a unique category, this method groups systems according to shared attributes, reflecting the fact that many deepfake pipelines reuse common components. 
By focusing on these attributes, models can more effectively trace spoofing algorithms that were unseen during training but are built from known elements during training \cite{klein24_interspeech}. Drawing on well‐defined properties from traditional text-to-speech (TTS) and voice conversion (VC) architectures, researchers have concentrated on classifying attributes such as input types \cite{klein24_interspeech}, acoustic models \cite{Zhang2023DistinguishingNS, klein24_interspeech}, speaker representations \cite{zhu2022source}, and vocoders \cite{zhu2022source, Yan2022AnII, Zhang2023DistinguishingNS, klein24_interspeech}, etc.
 
However, these prior studies target source tracing in traditional TTS/VC systems, and only initial efforts have explored codec‐based speech generation (CoSG) \cite{arora2025landscape} models. Xie et al.~\cite{xie2025neural} treat source tracing as a closed‐set classification problem—identifying known CoSG system IDs and using a binary out‐of‐distribution detector, but this approach fails when multiple unseen generators are encountered. 
Although there has been a series of recent works on source tracing \cite{mishra2025towards, negroni2025source, xiao2025listen, phukan2025towards}, few focus on codec-based deepfake speech \cite{wu24p_interspeech, chen2025codecfake+}.
More recently, Chen et al. \cite{chen2025codec} utilize the organized \textit{CodecFake+}\cite{chen2025codecfake+} taxonomy, which covers vector quantization schemes, auxiliary objectives, and decoder types, to formulate three source‐tracing tasks grounded in codec attributes. 
They \cite{chen2025codec} also show that a model trained on codec‐resynthesized data suffers from severe generalization issues when evaluated on CoSG‐generated speech, driven by unseen codec types. 

This paper further demonstrates that a model trained on codec re-synthesis speech tends to overfit to non-speech regions and unseen speech content. 
To address these challenges, we propose a Semantic-Acoustic Source Tracing Network (SASTNet) that leverages a semantic Encoder, a coarse-to-fine acoustic Encoder that jointly preserves content and captures fine-grained codec signatures, resulting in significant improvements in codec-based deepfake source tracing on \textit{CodecFake+} \cite{chen2025codecfake+}. 
Codebase will be released to support future research \footnote{\url{https://github.com/ResponsibleGenAI/CodecFake-Source-Tracing}}. 

%% file: sections/2.background.tex
\section{Background}\label{sec:background}
Neural audio codecs \cite{guo2025recent, wu2024towards,arora2025landscape, wu2024codec, wu2024codec_slt24} leave  ``fingerprints'' across three axes: vector quantization, auxiliary objectives, and decoder type. Together, these axes span most codec designs and enable robust tracing of codec-based deepfakes. 
Building on the taxonomy, Chen et. al.~\cite{chen2025codec} define three multi-class classification tasks for CodecFake source tracing:
\begin{itemize}
    \item \textbf{Vector Quantization Classification (VQ Task)}: 
    Four-class classification: classifies a waveform as real or generated by a model with codec using multi-codebook (Mvq), single-codebook (Svq), or scalar quantization (Scq).
    \item \textbf{Auxiliary Objective Classification (AUX Task)}: 
    Four-class classification: classifies a waveform as either real or generated by a model with a codec using semantic distillation (Sem), disentanglement objectives (Disent), or no auxiliary objective (None).
    \item \textbf{Decoder Type Classification (DEC Task)}: 
    Three-class classification: classifies an input utterance as real or generated by a model with a codec employing a time- or frequency-domain decoder (denoted Time and Freq). 
\end{itemize}
Note that we focus solely on the source-tracing tasks above, not on traditional anti-spoofing.
\input{Figs/model}

%% file: Figs/model.tex

\begin{figure*}[t]
    \centering
    \includegraphics[width=2.0\columnwidth]{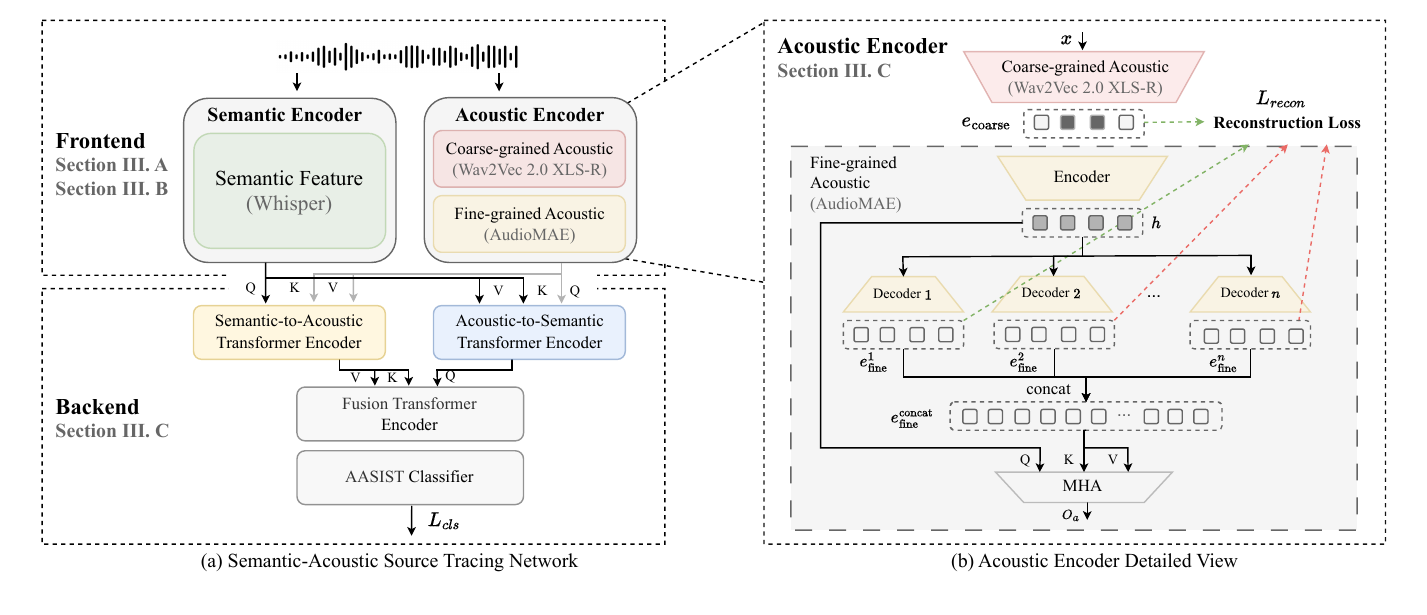}
    \vspace{-0.5em}
    \caption{The Overview of Semantic-Acoustic Source Tracing Network.}
    \label{fig:sastnet}
\end{figure*}

%% file: sections/3.SASTNet.tex
\section{Semantic-Acoustic Source Tracing Network}
In this section, we introduce our Semantic‐Acoustic Source Tracing Network (SASTNet) and explain the intuition behind its design. Chen et al. \cite{chen2025codec} previously applied a self-supervised learning model as a front-end feature extractor, followed by a classifier for source tracing, but their approach often emphasizes acoustic information at the expense of semantic cues. 
In contrast, we recognize that effective source tracing requires both the fine-grained acoustic patterns introduced by codec distortions and the preservation of semantic consistency. 
Indeed, Kawa et al. \cite{kawa23b_interspeech} have shown that combining semantic and acoustic features yields better capture of deepfake artifacts. 
Because relying solely on acoustic cues from codec distortions overlooks a critical fact: even codec-unprocessed speech exhibits spectral and temporal differences when the spoken content changes. 
If we focus only on acoustic features and ignore linguistic content, natural variations caused by different words or pronunciations can be mistaken for codec artifacts. 
In contrast, by preserving semantic information, the network can focus on the content-irrelevant information of the input utterance and then isolate only those subtle acoustic deviations that are independent of what is being said. 

Figure \ref{fig:sastnet}-(a) shows the framework of SASTNet. Given an input waveform, the model predicts the codec source label based on the tasks defined in Section \ref{sec:background}. SASTNet has three components: two front‐end encoders and a backend classifier. Front-end encoders include a Semantic Encoder, which extracts high‐level linguistic embeddings, and a Coarse‐to‐Fine Acoustic Encoder, which captures both coarse-grained and fine-grained codec‐specific acoustic features. The backend classifier is a component of three two‐layer cross‐modal Transformers, Semantic‐to‐Acoustic (SA), Acoustic‐to‐Semantic (AS), and Fusion Transformer, each built around a cross‐attention mechanism. In the SA Transformer, semantic features serve as queries while acoustic features act as keys and values; the AS Transformer swaps these roles. The Fusion Transformer then takes AS outputs as queries and SA outputs as keys and values to integrate both modalities. Finally, an AASIST \cite{jung2022aasist} backend model uses the fused representation to produce the source‐tracing prediction. The detailed settings of each block are described in the following sections.

\subsection{Semantic Encoder} 

The Semantic Encoder leverages a frozen, pre‐trained Whisper \cite{radford2022whisper} model to generate content‐robust embeddings that serve as a stable anchor for source tracing. By verifying that two utterances share identical linguistic content, these embeddings ensure the network ignores natural acoustic variations caused by different words or pronunciations and focuses only on codec‐induced distortions. 
Concretely, let \({x}\) denote the raw input waveform. We first zero‐pad \({x}\) to a fixed duration of 30 seconds, yielding ${x}_{\text{pad}}$.
We then pass \({x}_{\text{pad}}\) through Whisper’s encoder to obtain a feature tensor 
${s}_{\text{pad}} \in \mathbb{R}^{1500 \times 768}.$
Because only the initial tokens cover the same time span as our acoustic features, we truncate \({s}_{\text{pad}}\) to the first 256 non‐padding time steps:
${s}_{\text{trunc}} \in \mathbb{R}^{256 \times 768}.$
To align with the acoustic sequence length, we apply one‐dimensional average pooling over every two consecutive time steps in \({s}_{\text{trunc}}\). This yields the final semantic representation
${O}_{\text{s}} \in \mathbb{R}^{128 \times 768}.$
During training, Whisper’s weights remain frozen to preserve high‐level phoneme‐ and word‐level consistency. This design forces the acoustic branch to align its features only when the semantic anchor \({O}_{\text{s}}\) indicates that spoken content is identical.

\subsection{Coarse-to-Fine Acoustic Encoder} 
The goal of the acoustic branch is to isolate low‐level codec fingerprints while filtering out irrelevant noise and background artifacts. We design a coarse‐to‐fine acoustic encoder: a fine‐tuned Wav2Vec 2.0 XLS-R (0.3B) \cite{babu2021xls} first captures broad codec‐introduced anomalies, and a multi-decoder AudioMAE module then masks and reconstructs those embeddings to suppress noise and amplify residual, codec‐specific fingerprints. Our multi-decoder AudioMAE is extended by a single decoder AudioMAE \cite{huang2022masked}.
For simplicity, we refer to Wav2Vec 2.0 XLS-R (0.3B) as W2V2 in the following section. 
The view of the coarse-to-fine Acoustic Encoder is shown in Figure \ref{fig:sastnet}-(b). 

In the first stage, a W2V2 model fine-tuned on the corresponding source-tracing task is employed to extract robust, coarse-grained representations associated with different codecs. 
The initiation of fine-tuning here aims to enable W2V2 to capture more neural codec characteristics. 
Given an input waveform \( x \), the W2V2 generates initial embeddings \( e_{\text{coarse}} \). 

In the second stage, these embeddings are subsequently refined by a multi-decoder AudioMAE framework, which is inspired by the bottleneck architecture can purify some irrelevant information \cite{he2021mae, wu2022adversarial, chen24p_interspeech}.
Within AudioMAE, the encoder's output \( h \) is simultaneously passed to all decoders, each of which produces its own class-specific output $\{ e^{i}_{\text{fine}} \}_{i=1}^N$, where $N$ denotes the number of classification classes. 
Then, we use a multi-decoder scheme: since each class-specific autoencoder reconstructs inputs from its own class more accurately than others, the decoder with the best reconstruction indicates the corresponding class \cite{du2024towards}. 
Leveraging this property, we employ a multi-head attention (MHA) mechanism to enable the model to focus on the reconstruction quality across all decoders. The informative representation \( h \) serves as the query (Q), while \( e^{\text{concat}}_{\text{fine}} \), representing the concatenation of all reconstructed outputs from each decoder, serves as both the keys (K) and values (V).
The MHA output \( O_\text{a} \) serves as the final acoustic feature, allowing a more accurate identification of the codec responsible for generating the input audio. 

\subsection{Semantic‐Acoustic Fusion} 

The fusion module uses content‐robust semantic embeddings as an anchor with acoustic features, which capture inherent codec‐related patterns. 
Inspired by audio-visual learning \cite{chen2022push, chen2024mtd, chen2025localizingaudiovisualdeepfakeshierarchical}, a learnable cross‐attention mechanism \cite{vaswani2017attention} then extracts semantic–acoustic mismatch features, which are inspired by audio-visual synchronization. 
Specifically, three modules learn these mismatches from different aspects. 
Firstly, in the SA Transformer Encoder, semantic queries attend to acoustic keys/values to produce $O_{\text{SA}}$, forcing the semantic stream to highlight codec‐induced distortions. 
Secondly, in the AS Transformer Encoder, acoustic queries attend to semantic keys/values to produce $O_{\text{AS}}$, down-weighting any acoustic components aligned with the anchored content. 
Finally, these two conditioned maps are passed through a Fusion Transformer that uses $O_{\text{AS}}$ as queries and \(O_{\text{SA}}\) as keys and values to yield \(O_{\text{fusion}}\), which is then fed into the AASIST classifier. 

\subsection{Loss Function}
The \textit{SASTNet} integrates two loss functions: a classification loss $L_{\text{cls}}$ from the AASIST backend and a reconstruction loss $L_{\text{recon}}$ from AudioMAE. To balance these objectives, we adopt the Automatic Weighted Loss (AWL)~\cite{liebel2018auxiliary}, which computes a weighted sum of the reconstruction and classification losses:
\begin{equation}
L_{\text{overall}} = \text{AWL}(L_{\text{cls}},\;L_{\text{recon}})
\end{equation}
where two learnable weights,  $w_{\text{cls}}$ and $w_{\text{recon}}$, are automatically adjusted during training to optimize the contribution of each loss term.
The classification loss $L_{\text{cls}}$ is computed as the cross-entropy loss between the output scores of AASIST and the ground truth. 
The reconstruction loss $L_{\text{recon}}$ for multi-decoder AudioMAE is defined as follows:
\begin{equation}
    L_{\text{recon}} = L_{MSE_{y}} + \max\left(0, m + L_{MSE_{y}} - L_{MSE_{\text{other}}}\right)
\end{equation}
where $y$ denotes the input label, $m$ denotes the $margin$, and $L_{MSE_{y}}$ represents the Mean Squared Error between the masked region of the input \( e_{\text{coarse}} \) and the corresponding reconstructed region of output \( e^y_{\text{fine}} \) from  \( decoder_y \). 
$L_{MSE_{\text{other}}}$ refers to the average MSE between the masked region of \( e_{\text{coarse}} \) and corresponding reconstructed masked regions of other outputs \( e^i_{\text{fine}} \) from \( decoder_i \), for \( i \neq y \). Note that the input \( e_{\text{coarse}} \) is masked only during training. During inference, \( e_{\text{coarse}} \) is fed directly into the encoder without masking.

The goal is for the class-specific decoder to yield a lower reconstruction error compared to decoders not associated with the input class. This design ensures that each decoder is specialized in reconstructing inputs from its designated class. When the $L_{MSE_{\text{other}}}$ is lower than $L_{MSE_{y}}$, the second term in the loss becomes positive, increasing the overall reconstruction loss as a penalty. Conversely, if $L_{MSE_{y}}$ is smaller than $L_{MSE_{\text{other}}}$ by at least $m$ margin, the penalty term becomes zero, resulting in no additional loss. The margin $m = 0.1$ is set from observations that the MSE of the original single-decoder version of AudioMAE typically converges to 0.02–0.05. This setting enforces a minimum separation between the target decoder and the others, promoting class-specific specialization.

%% file: sections/4.setup.tex
\section{Source Tracing}
\subsection{Experimental Setup}
\label{sec:exp_setup}

We investigate source tracing in \textit{CodecFake+} \cite{chen2025codecfake+} dataset, which comprises two subsets: CoRS (speech resynthesized by pre-trained neural audio codec models) and CoSG (speech generated by codec-based speech generation models).  
For model training, we followed Chen et al. \cite{chen2025codec} and used taxonomy-guided balanced sampling to maximize data diversity under limited computational resources, selecting a subset of spoofed samples accordingly. The CoRS Train dataset contains 42,965 bona fide and 42,965 spoofed audio samples. Depending on the sampling strategy, different CoRS subsets are used. For example, when performing the AUX task, we use only the AUX subset of CoRS Train. 
For evaluation, we adopt the \textit{CodecFake+} protocol, which includes three subsets: CoRS Test, CoSG Test (known codec), and CoSG Test (All). 

Additionally, to assess the effects of unseen speech and silence on performance, we sampled subsets from CoRS: For training, we only use CoRS samples with utterance IDs \( \leq 250 \), referred to as the \textbf{CoRS Train (utt. ID \( \leq 250 \)) set}. For testing, we split into two subset: a \textbf{CoRS Test (Seen) set} with utterance IDs \( \leq 250 \) and \textbf{CoRS Test (Unseen) set} \( > 250 \).
During training, the model sees only utterances from CoRS Train (utt. IDs $\leq$ 250), while CoRS Test (Unseen) evaluates its ability to generalize to new content. 
For a detailed discussion, please refer to Section \ref{sec:reason_fail}.

We adopt W2V2-AASIST \cite{tak2022automatic} \footnote{\href{https://github.com/TakHemlata/SSL_Anti-spoofing}{https://github.com/TakHemlata/SSL\_Anti-spoofing}} as backbone and train three baseline models on the CoRS Train set under different sampling strategies, referred to as S-VQ, S-AUX, and S-DEC. 
All input waveforms with a length of 82{,}200 samples at a sampling rate of 16{,}000~Hz, with RawBoost~\cite{tak2022rawboost} augmentation. 
For AudioMAE, a patch size of 16 is used, and during training, a masking ratio of 0.4 is applied, divided equally between time and spectral dimensions. 
No masking is applied during inference. 
We train models on an NVIDIA RTX 4090 GPU with a batch size of 12 for 40 epochs. We use the Cosine Annealing learning rate (LR) scheduler with the initial learning rate is $5\times10^{-6}$, and the weight decay is $1\times10^{-4}$. 

%% file: sections/5.main_results.tex
\input{Tables/Tab_exp_vq_aux_dec_F1_only}

\subsection{Results}\label{sec:main_result}

We present a comparison between the performance of SASTNet and baseline models across three source tracing tasks in Table~\ref{tab:exp_result_vq_aux_dec}. 
Each row represents a distinct model, and corresponding F1 scores are derived from the model trained on the VQ, AUX, or DEC tasks. 
Each column denotes a different evaluation subset. 
Since this work focuses on improving single-task performance, we do not include multitask results for comparison with the previous work~\cite{chen2025codec}.

By examining F1‐score changes from CoRS to CoSG (kn. codec), we see that SASTNet outperforms all baselines on the three tasks but still experiences roughly a 50\% drop in F1‐score. This shows that, while more robust than prior models, SASTNet remains affected by artifacts introduced during generative modeling.
In CoSG (All), SASTNet’s performance in the AUX and DEC tasks drops by only 5\%–8\% compared to CoSG (kn. codec). Conversely, in the VQ task, SASTNet actually improves by 1.23\% on CoSG (All) versus CoSG (kn. codec). Taken together, these results demonstrate that SASTNet generalizes well to unseen codecs and maintains strong source tracing performance across different generative modeling and codec scenarios.

%% file: Tables/Tab_exp_vq_aux_dec_F1_only.tex
\begin{table}[t]
\centering
\fontsize{7}{9}\selectfont
\setlength\tabcolsep{4pt}
\caption{F1 Score (\%) $\uparrow$ of evaluation on different source tracing tasks.}
\label{tab:exp_result_vq_aux_dec}
\vspace{-1em}
\begin{minipage}{0.33\textwidth}
\centering
\textbf{(a) Vector Quantization Classification}
\vspace{0.5em}
\begin{tabularx}{\textwidth}{@{}cccc@{}}
\toprule
Model & \quad CoRS \quad & CoSG (kn. codec) & CoSG (All) \\
\midrule
Random  & 32.26      & 29.21          & 30.79         \\
S-VQ~\cite{chen2025codec}  & 97.39   & 42.94    & 35.26      \\
SASTNet  & \textbf{98.94} & \textbf{46.94} & \textbf{48.17}  \\
\bottomrule
\end{tabularx}
\end{minipage}%

\begin{minipage}{0.32\textwidth}
\centering
\textbf{(b) Auxiliary Objective Classification}
\vspace{0.5em}
\begin{tabularx}{\textwidth}{@{}cccc@{}}
\toprule
Model & CoRS & CoSG (kn. codec) & CoSG (All) \\
\midrule
Random   & 28.78   & 30.94          & 31.04         \\
S-AUX~\cite{chen2025codec} & 97.15   & 32.80  & 19.44  \\
SASTNet   & \textbf{97.90} & \textbf{48.13} & \textbf{43.47} \\
\bottomrule
\end{tabularx}
\end{minipage}%

\begin{minipage}{0.32\textwidth}
\centering
\textbf{(c) Decoder Types Classification}
\vspace{0.5em}
\begin{tabularx}{\textwidth}{@{}cccc@{}}
\toprule
Model & CoRS & CoSG (kn. codec) & CoSG (All) \\
\midrule
Random & 38.97      & 36.89      & 34.37  \\
S-DEC~\cite{chen2025codec}  & \textbf{97.96}   & 46.73  & 27.26  \\
SASTNet   & 97.72    & \textbf{48.03}    & \textbf{40.71}     \\
\bottomrule
\end{tabularx}
\end{minipage}
\vspace{-3mm}
\end{table}

%% file: sections/6.analysis_and_discussion.tex
\input{Figs/png_asru/failure_cases}
\input{Tables/Table_exp_cors_unseen_speech}

\section{Analysis and discussion}
\subsection{The Reason for Model Generalization Failures}
\label{sec:reason_fail}

Prior work \cite{chen2025codec} discussed how unseen codec types and generative modeling may affect model 
generalization. In this section, we further identify two additional factors, silence and unseen content, that affect model generalization.

\subsubsection{Silence}
One challenge is silence variability: inconsistent pauses at utterance boundaries disrupt codec-signature alignment, causing models to focus on noise in silent segments rather than learning content-independent features. Figure \ref{fig:failures_cases}(a) shows that CoRS contains silent intervals in both bona fide and resynthesized speech.
This occurs because the CoRS samples in \textit{CodecFake+} are derived from the untrimmed version of VCTK \cite{yamagishi2019cstr}, which includes silence at the beginning and end of each utterance. Since silence is present in both bona fide and spoofed samples of different spoofing algorithms, it may go unnoticed during lab experiments, but later leads to degradation when encountering samples with different silence characteristics in the real world. 
When the model is trained on CoRS with silence at the beginning and end of each utterance, it can achieve perfect source-tracing F1 scores on samples with long silences as shown in Figure \ref{fig:failures_cases}(b). 
However, the model suffers performance degradation when facing samples with short silences, e.g., from UniAudio \cite{pmlr-v235-yang24x} and MaskGCT \cite{wang2024maskgct}. We further analyzed the correlation between silence and performance by correlating the proportion of silent samples with F1 scores in CoSG ($\rho_{\text{VQ}}=0.62$, $\rho_{\text{AUX}}=0.60$, $\rho_{\text{DEC}}=0.75$). These findings also appear in deepfake detection \cite{zhang2023impact, zhang122021effect} and highlight the need to model robustly against silence variability. 

\subsubsection{Unseen Content or Speakers}
Unseen content and speakers are both notable challenges, which means that linguistic contexts or speaker information not present in the training set introduce phonetic and prosodic patterns outside the model’s learned distribution, thereby reducing performance. 
\noindent\textbf{Experiments:} We conducted a controlled experiment on CoRS set to isolate the effects of silence and variability in speech content on source-tracing performance. During training, the model was trained on the \textbf{CoRS Train (utt. IDs $\leq$250) set}. 
For evaluation, we used \textbf{CoRS Test (Seen) set} and \textbf{CoRS Test (Unseen) set} to assess the performance of the model in both familiar and unfamiliar speech content.
Additionally, we evaluated the impact of silence by comparing model performance on audio with and without silent segments, aiming to determine whether the model had overfitted to silence. 

\noindent\textbf{Results:} Table~\ref{tab:BASELINE_CORS_UNSEEN} presents the results of single-task baseline models evaluated on the CoRS Test (Unseen) set, along with the CoRS Test (Seen) set. Excluding silent segments from the input led to a substantial performance decline, with the single-task model’s F1 score decreasing by approximately 27\%. This suggests that the presence of silence has a detrimental effect on the baseline models’ performance. When evaluated on seen speech, the F1 score is approximately 69\%, whereas on unseen speech, it decreases to around 67\%, indicating that the model’s effectiveness further diminishes when processing unseen speech content.
This suggests that the CoSG dataset, which lacks silence and unseen speech components relative to CoRS, poses a challenge for models to generalize to CoSG.

In summary, in Sections \ref{sec:main_result} and \ref{sec:reason_fail}, we show how four variables, namely generative modeling process, unseen codec types, silence diversity, and unseen speech, combine to limit model generalization and tracing fidelity.

\input{Tables/Table_preliminary_ablation_silence}

\subsection{The Impact of Semantic and Acoustic Encoder} 
To gain deeper insights into how the semantic and acoustic encoders contribute to generalization under the effects of silence, unseen speech, codec type, and generative modeling, we conducted an ablation study on the AUX task and present the results in Tables \ref{tab:ablation_sas_sil} and \ref{tab:ablation_encoder}. We compare models using different semantic and acoustic encoders by evaluating various front‐end configurations as follows: 
\begin{itemize}
    \item \textbf{MAE-only}: This configuration uses mel-spectrograms as input to the MAE module. We further divide this setting into \texttt{S-MAE} (Single-Decoder MAE) and \texttt{M-MAE} (Multi-Decoder MAE) based on the number of decoders used.
    \item \textbf{Whisper + MAE}: The MAE processes mel-spectrograms to extract acoustic features, which are then fused with semantic features extracted from Whisper.

    \item \textbf{W2V2 + MAE}: Here, we replace the MAE’s mel‐spectrogram input with W2V2 SSL features. We consider two variants: \texttt{Pretrained W2V2 + M-MAE}, where W2V2 is initialized with pretrained weights; and \texttt{Tuned W2V2 + M-MAE}, where W2V2 is initialized with the S-AUX task-specific weights.

    \item \textbf{Our proposed SASTNet} has two variants: \texttt{SASTNet (Pretrained W2V2)}, where W2V2 is initialized with pretrained weights; and \texttt{SASTNet (Tuned W2V2)}, where W2V2 is initialized with the \texttt{S-AUX} weights.
\end{itemize}

\subsubsection{Effects of Silence and Unseen Speech.}
In Table~\ref{tab:ablation_sas_sil}, 
compared to \texttt{S-AUX}, using MAE alone as the front-end resulted in a significant decline in F1 score due to the removal of silence, with \texttt{S-MAE} and \texttt{M-MAE} showing drops of 49.69 and 37.36, respectively. 
Notably, \texttt{M-MAE} demonstrated greater robustness to silence removal than \texttt{S-MAE}. 
The \texttt{Whisper+MAE} exhibited a smaller F1 drop of 23.87, indicating higher robustness compared to both \texttt{S-AUX} and the MAE-only settings. 
Similarly, \texttt{W2V2+MAE} showed improved robustness relative to MAE-only, with \texttt{Pretrained W2V2+M-MAE} and \texttt{Tuned W2V2+M-MAE} resulting in drops of 16.46 and 14.64, respectively. 
This suggests that initializing \texttt{W2V2} with \texttt{S-AUX} improves robustness to silence. Finally, \texttt{SASTNet (Pretrained W2V2)} and \texttt{SASTNet (Tuned W2V2)} exhibited the smallest performance degradation, with drops of 9.07 and 8.75, respectively—approximately half the drop observed with \texttt{W2V2+MAE}.

\input{Tables/Table_impact_of_encoder}
For unseen speech content without silence, F1 scores improve across models, from 39.27 for \texttt{S-MAE} to 89.76 for \texttt{SASTNet (Tuned W2V2)}, indicating that multi-decoder reconstruction, Fine-tuned W2V2, and Whisper-based features each play a crucial role in enhancing robustness under silence removal and unseen speech conditions.
Moreover, mel-spectrogram inputs exhibit pronounced sensitivity to speech content: when the input of MAE is mel-spectrograms, the F1 gap between seen and unseen speech is considerable. In contrast, substituting mel-spectrograms with SSL features markedly reduces this discrepancy, indicating that SSL features are more invariant to content than mel-spectrograms. 

\input{Figs/figs_Confusion_Matrix}

\subsubsection{Effects of Codec Type and Generative Modeling.}
In Table~\ref{tab:ablation_encoder}, the performance of \texttt{S-AUX} on CoSG is notably poor, even falling below the F1 score of the random baseline. This indicates that \texttt{S-AUX} fails to perform source tracing on CoSG data. In contrast, using MAE as the front-end yields better results. Specifically, on CoSG (All), \texttt{S-MAE} achieves a 35.59 F1 score, while \texttt{M-MAE} reaches 36.64. Across the 17 CoSG models, \texttt{M-MAE} consistently outperforms \texttt{S-MAE} in three unseen codecs and in 9 out of 14 seen codecs. These results suggest that multi-decoder reconstruction improves robustness when handling unseen codecs and generative modeling.

Moreover, (\texttt{Whisper+MAE}) integrating Whisper with MAE leads to a slight performance improvement over MAE alone, achieving F1 scores of 41.74 on CoSG (kn.) and 36.94 on CoSG (All). Similarly, incorporating W2V2 with MAE (\texttt{W2V2+MAE}) also enhances performance. Specifically, \texttt{Pretrained W2V2+M-MAE} yields F1 scores of 43.29 on CoSG (kn.) and 36.89 on CoSG (All), while \texttt{Tuned W2V2+M-MAE} achieves 46.18 and 41.91, respectively. These findings indicate that SSL-based features are more robust than traditional mel-spectrograms in handling unseen codecs and generative modeling variations. The superior performance of the \texttt{Tuned W2V2} could be attributed to its prior knowledge of codec-related data, whereas the \texttt{Pretrained W2V2} was only pre-trained on bona fide samples. As a result, the \texttt{Tuned W2V2} is better equipped to capture codec-specific artifacts.

Finally, \texttt{SASTNet (Pretrained W2V2)} achieves F1 scores of 47.28 on CoSG (kn.) and 40.62 on CoSG (All), while \texttt{SASTNet (Tuned W2V2)} achieves 48.13 and 43.47, respectively. These results represent a 2\~4\% F1 improvement over \texttt{W2V2+MAE}, highlighting the effectiveness of fusing refined SSL and Whisper features. This fusion enables the model to better focus on codec-related artifacts with semantic anchors, thereby enhancing performance on CoSG data.

\subsection{Discussion}
To better understand results, we compared the confusion matrices of SASTNet and the baseline on the CoSG (All) dataset, with the results shown in Fig.~\ref{fig:cm_compare}. Notably, SASTNet demonstrates significantly improved accuracy in detecting unseen real speech across the VQ, AUX, and DEC source tracing tasks. Specifically, on CoSG (All), the accuracy for unseen real speech improves by 2.63$\times$, 2.7$\times$, and 3.36$\times$ compared to the baseline, respectively. 
Moreover, SASTNet shows improved performance in identifying MVQ, None, Disentanglement, and Frequency-domain codecs, suggesting that it generalizes better across various codec attributes. This can be attributed to its joint modeling of semantic and acoustic features, which allows the model to compare bona fide and codec-resynthesized versions of the same utterance. By leveraging consistent semantic content with contrast acoustic cues, SASTNet is better equipped to capture subtle artifacts left from codecs, thereby enhancing its detection performance.

In Figure \ref{fig:attention_map_comparison}, we visualize three cross-attention maps from the last layer of each transformer encoder for a randomly chosen bona fide Encoder synthesis pair.
Figures (a) and (d) show opposite highlighted regions, indicating that the SA Transformer focuses on semantic–acoustic mismatches information (codec distortions). For resynthesized audio, this produces a wider, more dispersed bright band.
Figures (b) and (e) remain flat and uniform for bona fide speech, and only exhibit slight brightness at distorted frames for spoofed audio. This shows that the AS module preserves content-aligned regions while suppressing interference.
Figures (c) and (f) combine both patterns in spoofed audio, with highlights at content-aligned and distorted information. 
This implies that the Fusion module merges both patterns so the classifier can focus on codec-specific fingerprints, regardless of content, thus improving detection precision and generalization. 

\input{Figs/figs_attention_map}

%% file: Figs/png_asru/failure_cases.tex

\begin{figure}[t]
    \centering
    \begin{minipage}[t]{0.48\textwidth} 
        \begin{minipage}[t]{0.49\textwidth}
            \centering
            \textbf{\small (a) CoRS}\par\vspace{0.3em}
            
            \includegraphics[width=\linewidth]{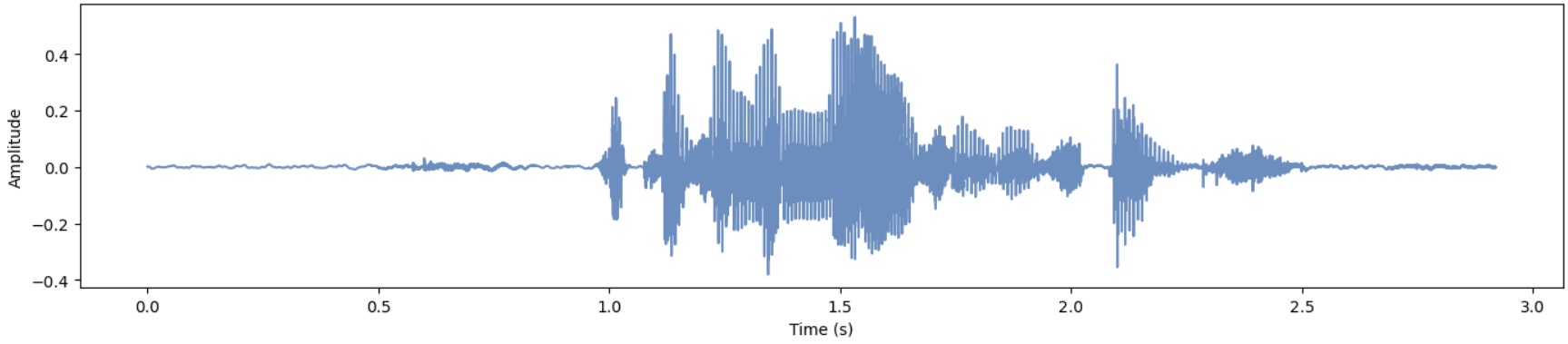}
            \scriptsize p227\_306 (real)\par\vspace{0.5em}
            
            \includegraphics[width=\linewidth]{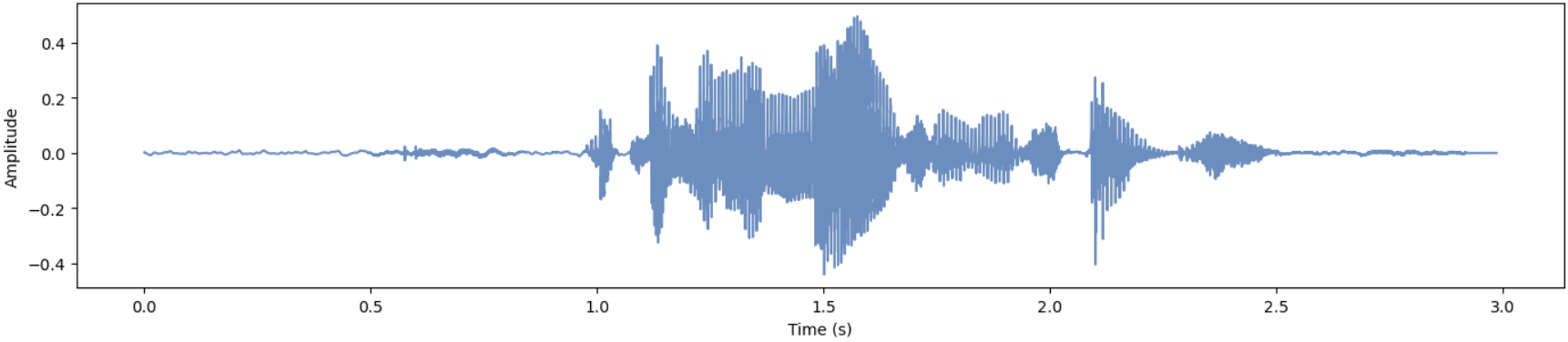}
            \scriptsize p227\_306 (snac)\par\vspace{0.5em}
            
            \includegraphics[width=\linewidth]{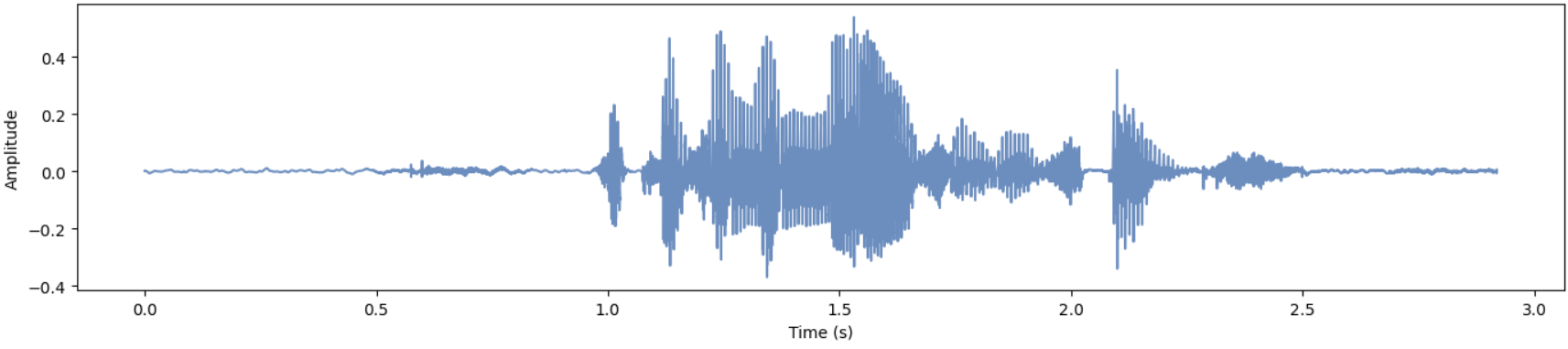}
            \scriptsize p227\_306 (DAC24)\par\vspace{0.3em}
        \end{minipage}
        \hfill
        \begin{minipage}[t]{0.49\textwidth}
            \centering
            \textbf{\small (b) CoSG}\par\vspace{0.3em}
            
            \includegraphics[width=\linewidth]{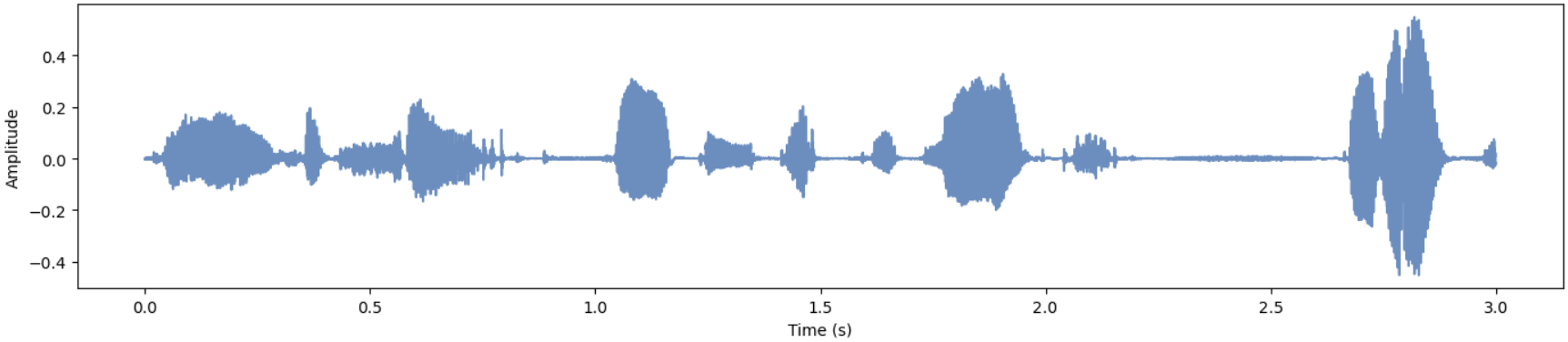}
            \scriptsize UniAudio (S-AUX\_F1=0\%)\par\vspace{0.5em}
            
            \includegraphics[width=\linewidth]{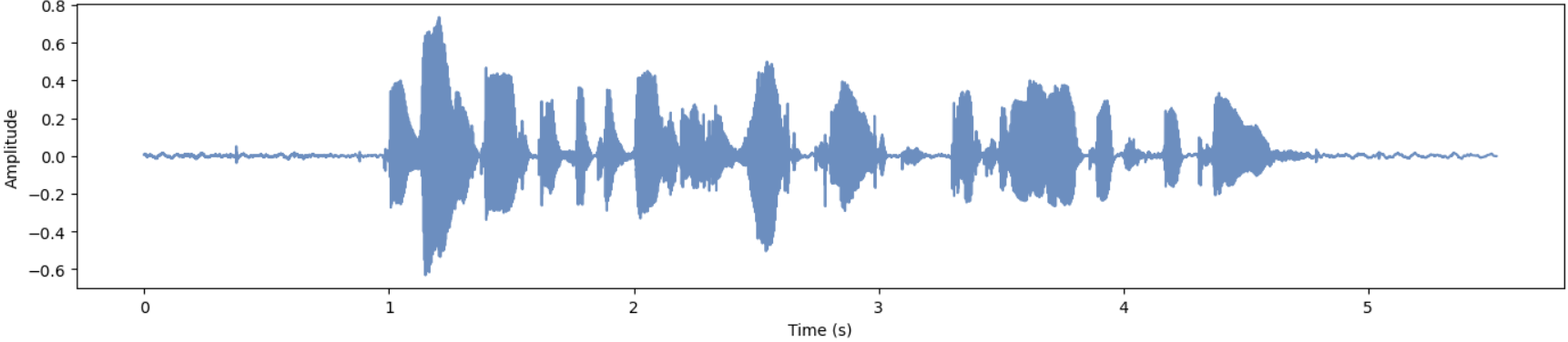}
            \scriptsize USLM (S-AUX\_F1=100\%)\par\vspace{0.5em}
            
            \includegraphics[width=\linewidth]{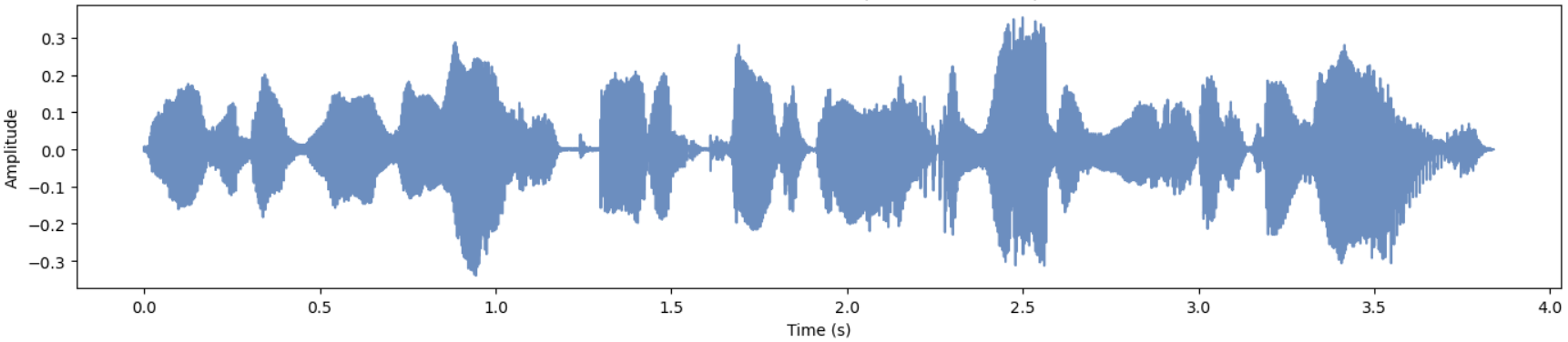}
            \scriptsize MaskGCT (S-AUX\_F1=5.76\%)\par\vspace{0.3em}
        \end{minipage}
    \end{minipage}
    \caption{The failure case study of codec-based deepfake source tracing.}
    \vspace{-3mm}
    \label{fig:failures_cases}
\end{figure}

%% file: Tables/Table_exp_cors_unseen_speech.tex
\begin{table}[t]
  \centering
  \fontsize{7}{9}\selectfont
  \setlength\tabcolsep{5pt}
  \caption{F1 Scores (\%) $\uparrow$ of Baseline Models Trained on CoRS Train (utt. IDs $\leq$ 250) and Tested on CoRS Test (SEEN \& Unseen)}
  \label{tab:BASELINE_CORS_UNSEEN}
  \vspace{-1em}
  \begin{tabularx}{0.44\textwidth}{@{} c c c c c c c @{}} 
    \toprule
    \multirow{2}{*}{Model} & \multicolumn{3}{c}{CoRS Test (Seen)} & \multicolumn{3}{c}{CoRS Test (Unseen)} \\
    \cmidrule(lr){2-4} \cmidrule(lr){5-7}
         & original data & rm\_sil & diff & original data & rm\_sil & diff \\
    \midrule
    S-VQ   & 93.62 & 69.04 & -24.58 & 93.98 & 67.55 & -26.43 \\
    S-AUX  & 93.70 & 69.45 & -24.25 & 94.18 & 66.93 & -27.25 \\
    S-DEC  & 92.05 & 68.21 & -23.84 & 93.09 & 67.52 & -25.57 \\
    \bottomrule
  \end{tabularx}
  \vspace{-3mm}
\end{table}

%% file: Tables/Table_preliminary_ablation_silence.tex
\begin{table}[t]
  \centering
  \scriptsize
  \caption{Ablation Analysis of Silence and Unseen Speech Effects on SASTNet.  
    Models were trained on CoRS Train (utt. ID$\leq$250) and evaluated on CoRS Test (Seen \& Unseen) using F1-score.}
  \label{tab:ablation_sas_sil}
  \resizebox{\columnwidth}{!}{%
  \begin{tabular}{
    >{\centering\arraybackslash}m{2.95cm}|
    >{\centering\arraybackslash}m{0.3cm}
    >{\centering\arraybackslash}m{0.3cm}
    >{\centering\arraybackslash}m{0.8cm}|
    cc|cc
  }
    \toprule
      \multirow{2}{*}{\textbf{Model}} 
      & \multicolumn{3}{c|}{\textbf{Front-end Modules}} 
      & \multicolumn{2}{c|}{\textbf{CoRS (Seen)}} 
      & \multicolumn{2}{c}{\textbf{CoRS (Unseen)}} \\
    \cmidrule(lr){2-4} \cmidrule(lr){5-6} \cmidrule(lr){7-8}
        &  W2V2  &  MAE  &  Whisper  & orig. & rm\_sil & orig. & rm\_sil \\
    \midrule
    S-AUX       & {\color{Green}\cmark} &     &     & 93.70 & 69.45 & 94.18 & 66.93 \\
    \midrule
    S-MAE \cite{wang2024genuine}       &     & {\color{Green}\cmark} &     & 89.08 & 65.27 & 88.96 & 39.27 \\
    M-MAE       &     & {\color{Green}\cmark} &     & 93.42 & 67.73 & 92.66 & 55.30 \\
    \midrule
    Whisper + M-MAE  &     & {\color{Green}\cmark} & {\color{Green}\cmark} & 85.01 & 71.45 & 84.16 & 60.29 \\
    \midrule
    Pretrained W2V2+ M-MAE   & {\color{Green}\cmark} & {\color{Green}\cmark} &     & 95.68 & 79.96 & 96.37 & 79.91 \\
    Tuned W2V2+ M-MAE        & {\color{Green}\cmark} & {\color{Green}\cmark} &     & 96.56 & 83.69 & 96.79 & 82.15 \\
    \midrule
    SASTNet (Pretrained W2V2) & {\color{Green}\cmark} & {\color{Green}\cmark} & {\color{Green}\cmark} & 95.10 & 86.54 & 95.29 & 86.22 \\
    SASTNet (Tuned W2V2)    & {\color{Green}\cmark} & {\color{Green}\cmark} & {\color{Green}\cmark} & \textbf{98.74} & \textbf{89.77} & \textbf{98.51} & \textbf{89.76} \\
    \bottomrule
  \end{tabular}} 
\end{table}

%% file: Tables/Table_impact_of_encoder.tex

\begin{table}[tb]
  \centering
  \scriptsize
  \caption{Ablation Study of Codec Types and Generative Modeling Effects on SASTNet. 
  The model was trained on the CoRS Train set using AUX balanced sampling and evaluated on the standard \textit{CodecFake+} evaluation set. 
  }
  \label{tab:ablation_encoder}
  \setlength{\tabcolsep}{3pt}
  \begin{tabular}{
  >{\centering\arraybackslash}m{3.0cm}|>{\centering\arraybackslash}m{0.63cm}>{\centering\arraybackslash}m{0.63cm}>{\centering\arraybackslash}m{0.84cm}|>{\centering\arraybackslash}m{0.63cm}>{\centering\arraybackslash}m{0.7cm}>{\centering\arraybackslash}m{0.63cm}}
    \toprule
      \multirow{2}{*}{\textbf{Model}} 
      & \multicolumn{3}{c|}{\textbf{Front-end Modules}} 
      & \multicolumn{3}{c}{\textbf{F1 (\%) $\uparrow$}} \\
    \cmidrule(lr){2-4} \cmidrule(lr){5-7}
    
        & \textbf{W2V2} & \textbf{MAE} & \textbf{Whisper} 
      & \textbf{CoRS} & \textbf{\makecell{CoSG\\(kn.)}} & \textbf{\makecell{CoSG\\(All)}} \\
    \midrule
     Random           &     &     &     & 28.78 & 30.94 & 31.04 \\
    \midrule
    S-AUX               & {\color{Green} \cmark} &     &     & 97.15 & 32.80 & 19.44 \\
    \midrule
    S-MAE \cite{wang2024genuine}            &     & {\color{Green} \cmark} &     & 96.67 & 42.21 & 35.59 \\
    M-MAE            &     & {\color{Green} \cmark} &     & 97.10 & 38.41 & 36.64 \\
    \midrule
    Whisper + M-MAE  &     & {\color{Green} \cmark} & {\color{Green} \cmark} & 95.98 & 41.74 & 36.94 \\
    \midrule
    Pretrained W2V2+ M-MAE  & {\color{Green} \cmark} & {\color{Green} \cmark} &     & 98.62 & 43.29 & 36.89 \\
    Tuned W2V2+ M-MAE        & {\color{Green} \cmark} & {\color{Green} \cmark} &     & 95.81 & 46.18 & 41.91 \\
    \midrule
    SASTNet (Pretrained W2V2) & {\color{Green} \cmark} & {\color{Green} \cmark} & {\color{Green} \cmark} & 94.80 & 47.28 & 40.62 \\
    SASTNet (Tuned W2V2)    & {\color{Green} \cmark} & {\color{Green} \cmark} & {\color{Green} \cmark} & 97.90 & 48.13& 43.47 \\
    \bottomrule
  \end{tabular}
\end{table}

%% file: Figs/figs_Confusion_Matrix.tex
\begin{figure}[t]
    \centering
    \subfloat[\footnotesize \textnormal{S-VQ}]{%
        \includegraphics[width=0.3 \linewidth]{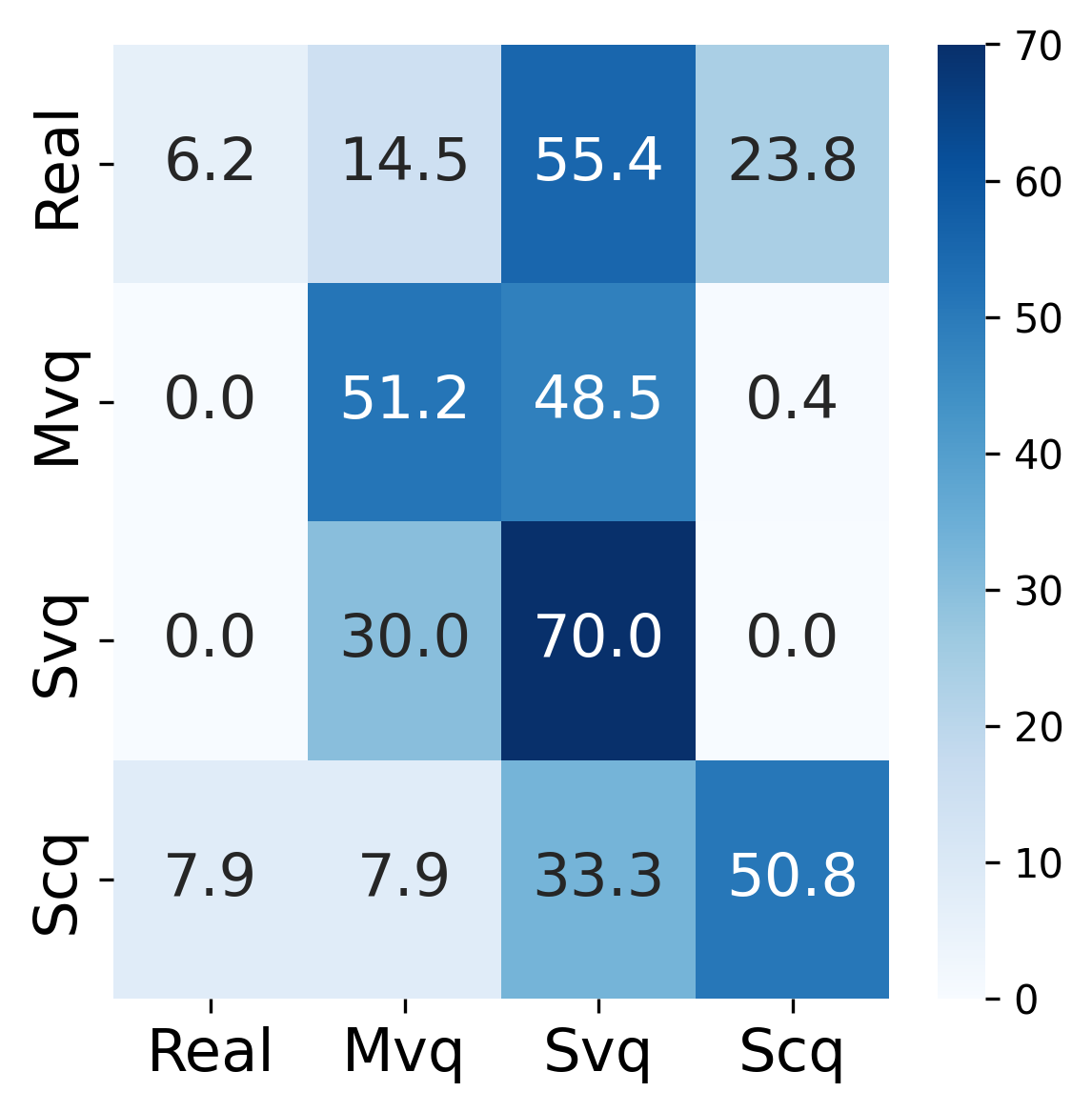}%
        \label{fig:cm_s1}
    }
    \hfill
    \subfloat[\footnotesize \textnormal{S-AUX}]{%
        \includegraphics[width=0.3 \linewidth]{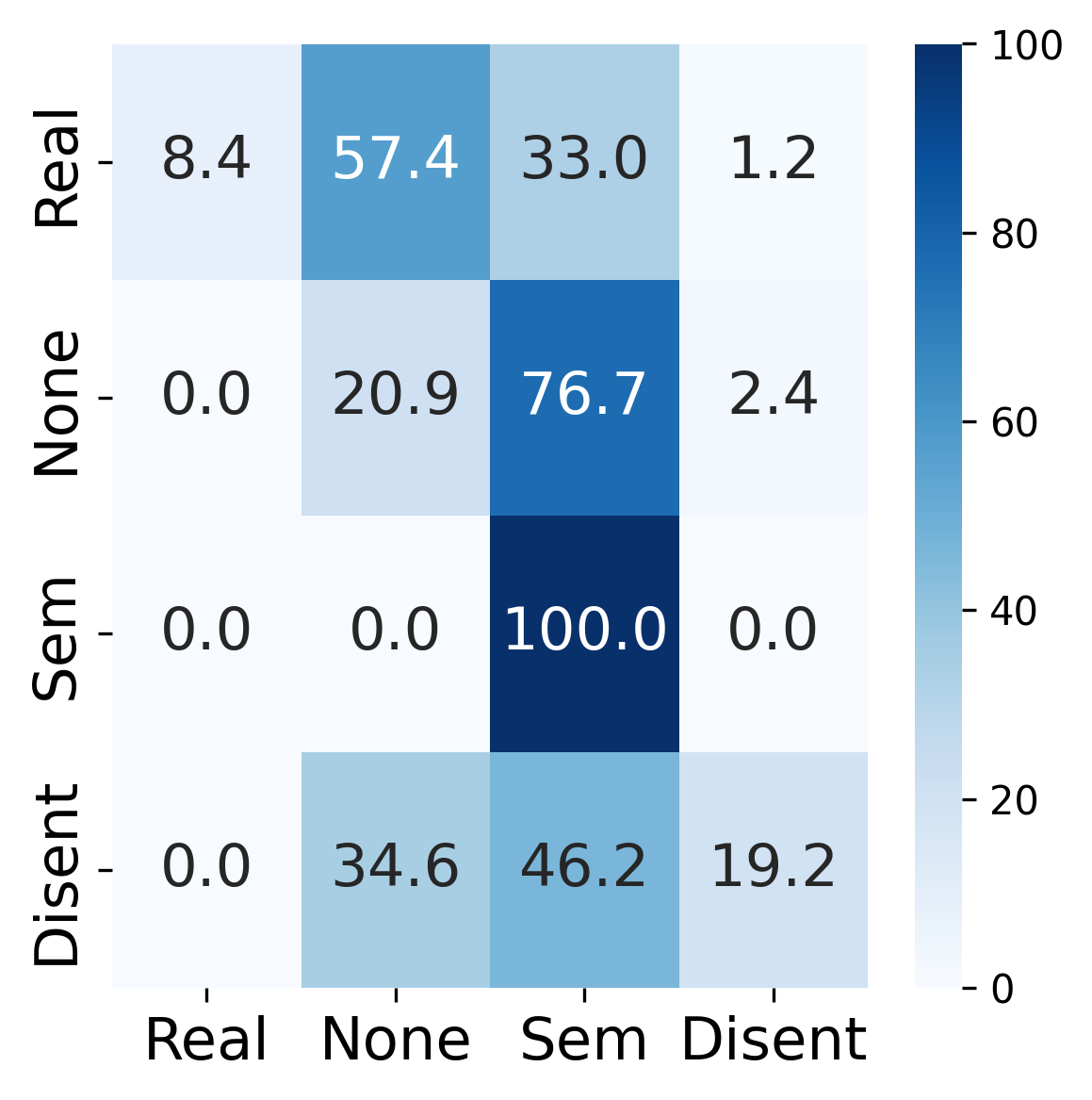}%
        \label{fig:cm_s2}
    }
    \hfill
    \subfloat[\footnotesize \textnormal{S-DEC}]{%
        \includegraphics[width=0.3 \linewidth]{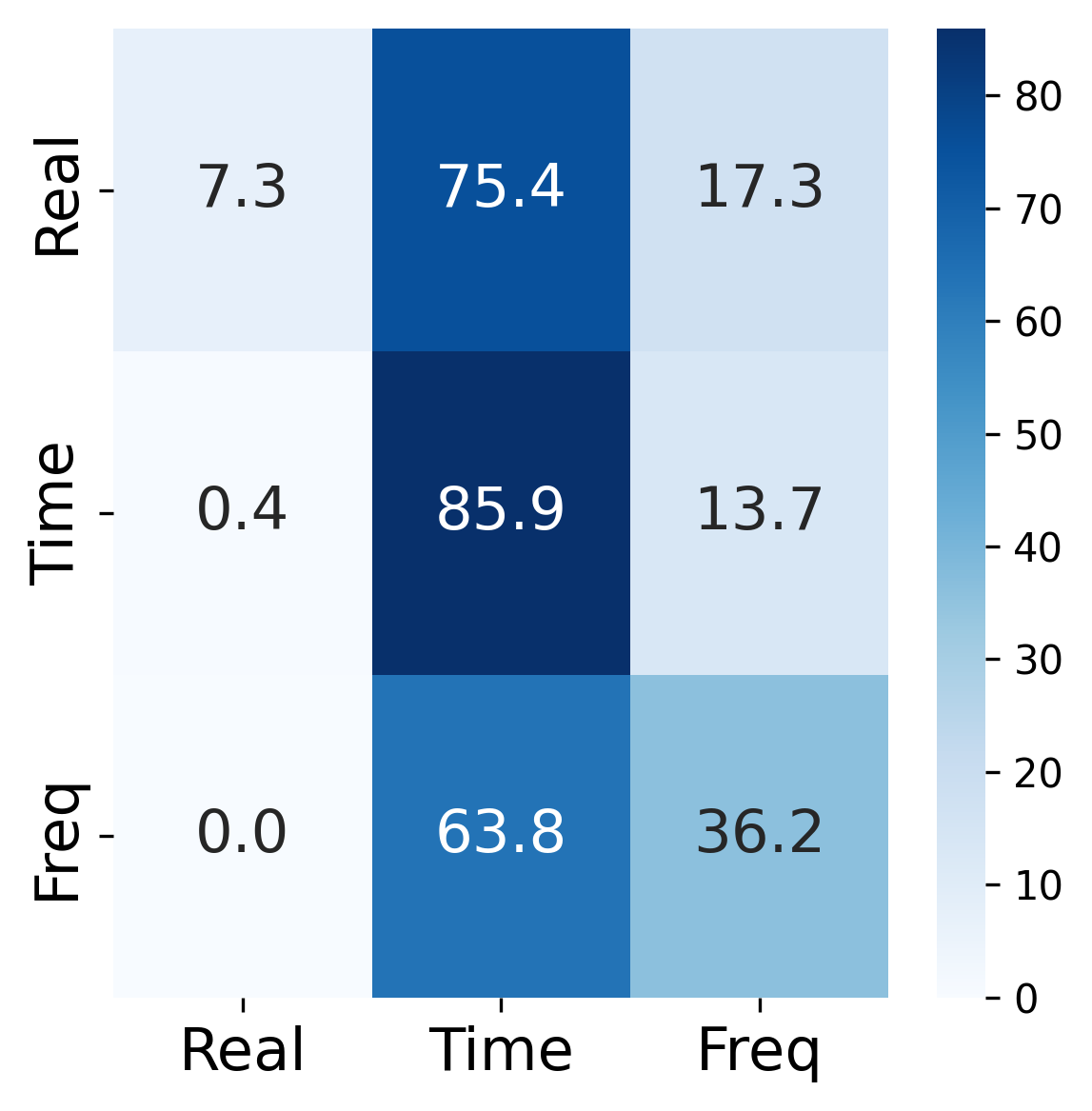}%
        \label{fig:cm_s3}
    }
    \vspace{-1em}
    \subfloat[\footnotesize \textnormal{SASTNet (VQ)}]{%
        \includegraphics[width=0.3 \linewidth]{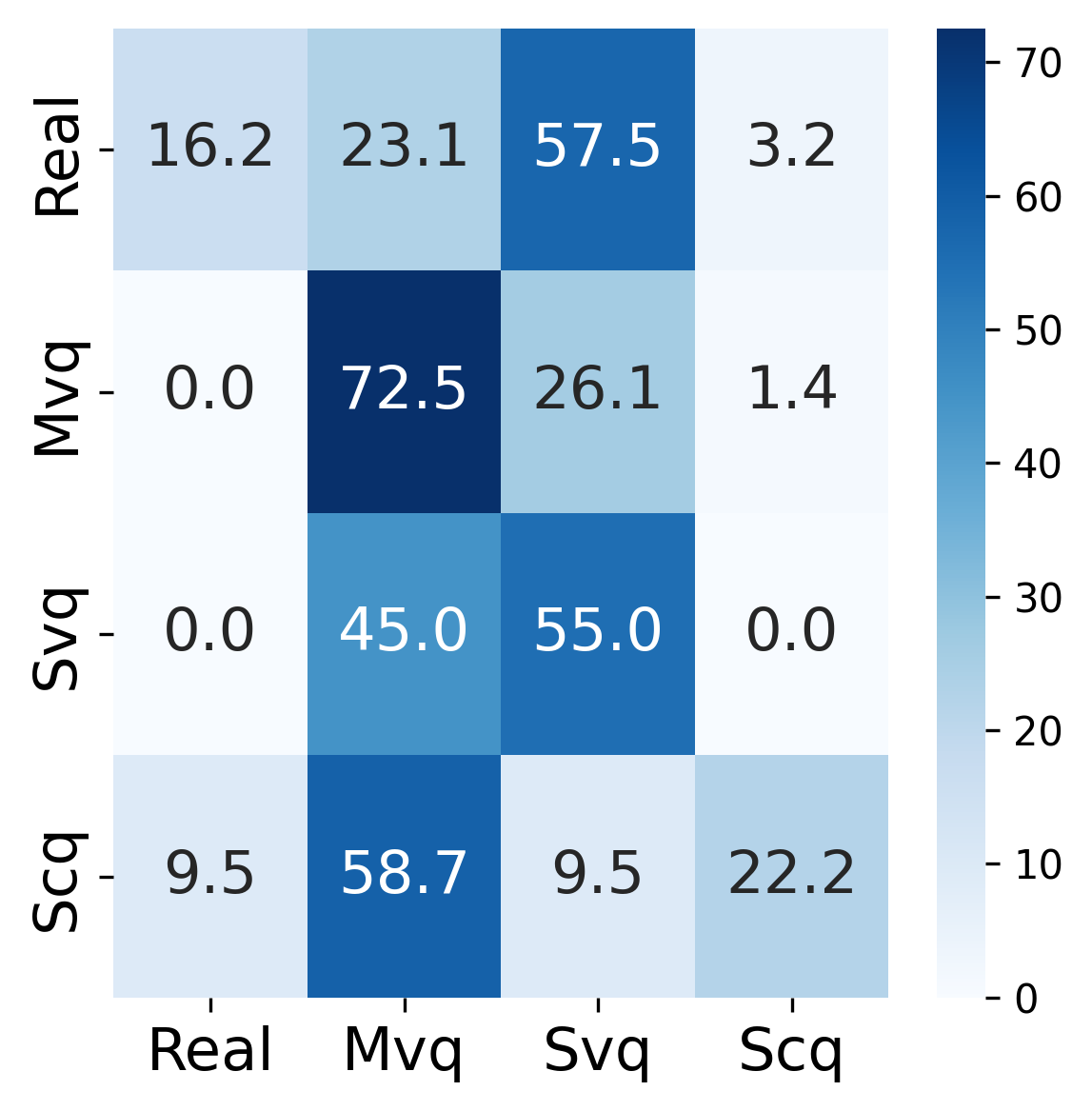}%
        \label{fig:cm_sast_vq}
    }
    \hfill
    \subfloat[\footnotesize \textnormal{SASTNet (AUX)}]{%
        \includegraphics[width=0.3 \linewidth]{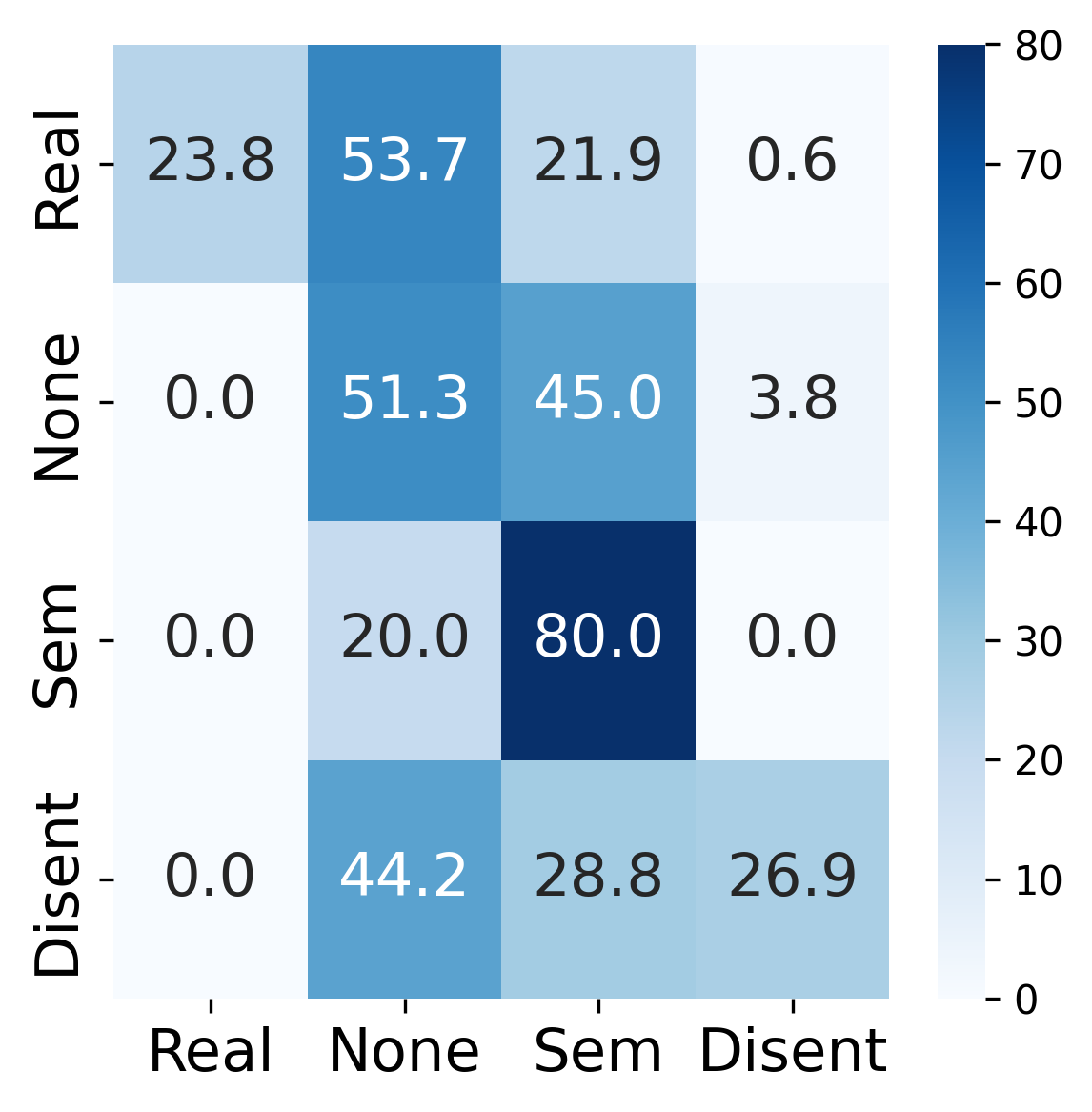}%
        \label{fig:cm_sast_aux}
    }
    \hfill
    \subfloat[\footnotesize \textnormal{SASTNet (DEC)}]{%
        \includegraphics[width=0.3 \linewidth]{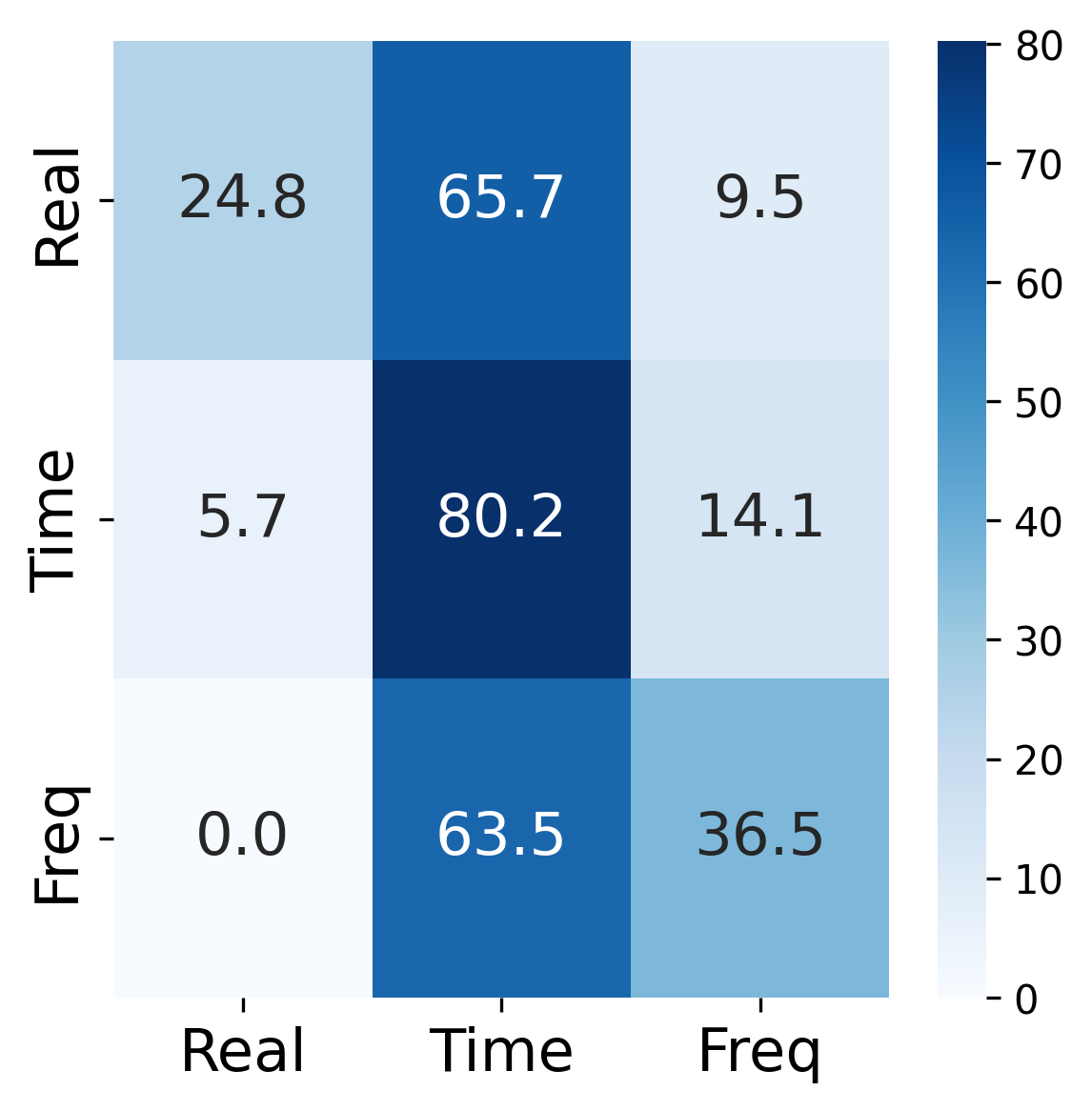}%
        \label{fig:cm_sast_dec}
    }
    \caption{Confusion matrices of baseline and SASTNet on CoSG (All) evaluation set, row-normalized by true labels, with predictions on the horizontal axis and true labels on the vertical axis. More details about labels in Section~\ref{sec:background}}
    \vspace{-3mm}
    \label{fig:cm_compare}
\end{figure}

%% file: Figs/figs_attention_map.tex
\begin{figure}[t]
    \centering
    \subfloat[\footnotesize \textnormal{Bona fide (S$\rightarrow$A)}]{%
        \includegraphics[width=0.3 \linewidth]{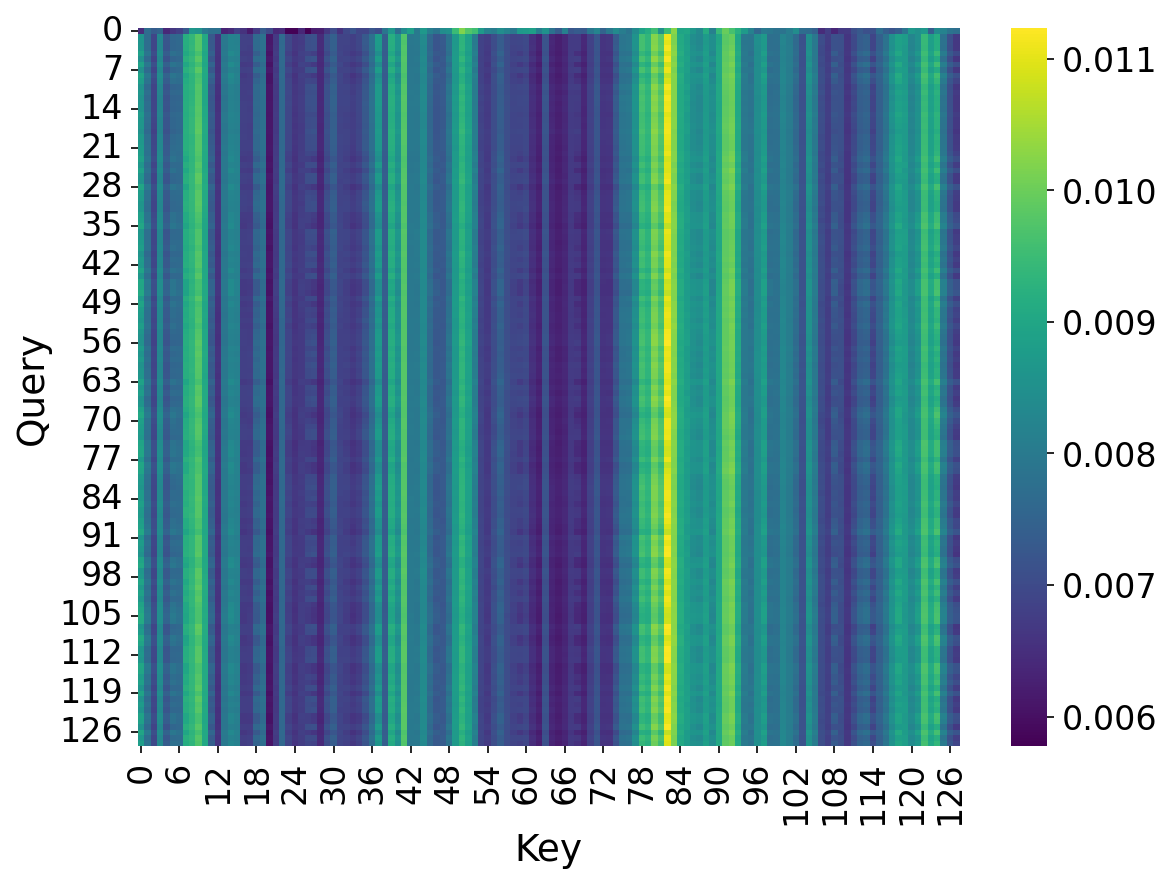}%
        \label{fig:att_sa_bf}
    }
    \hfill
    \subfloat[\footnotesize \textnormal{Bona fide (A$\rightarrow$S)}]{%
        \includegraphics[width=0.3 \linewidth]{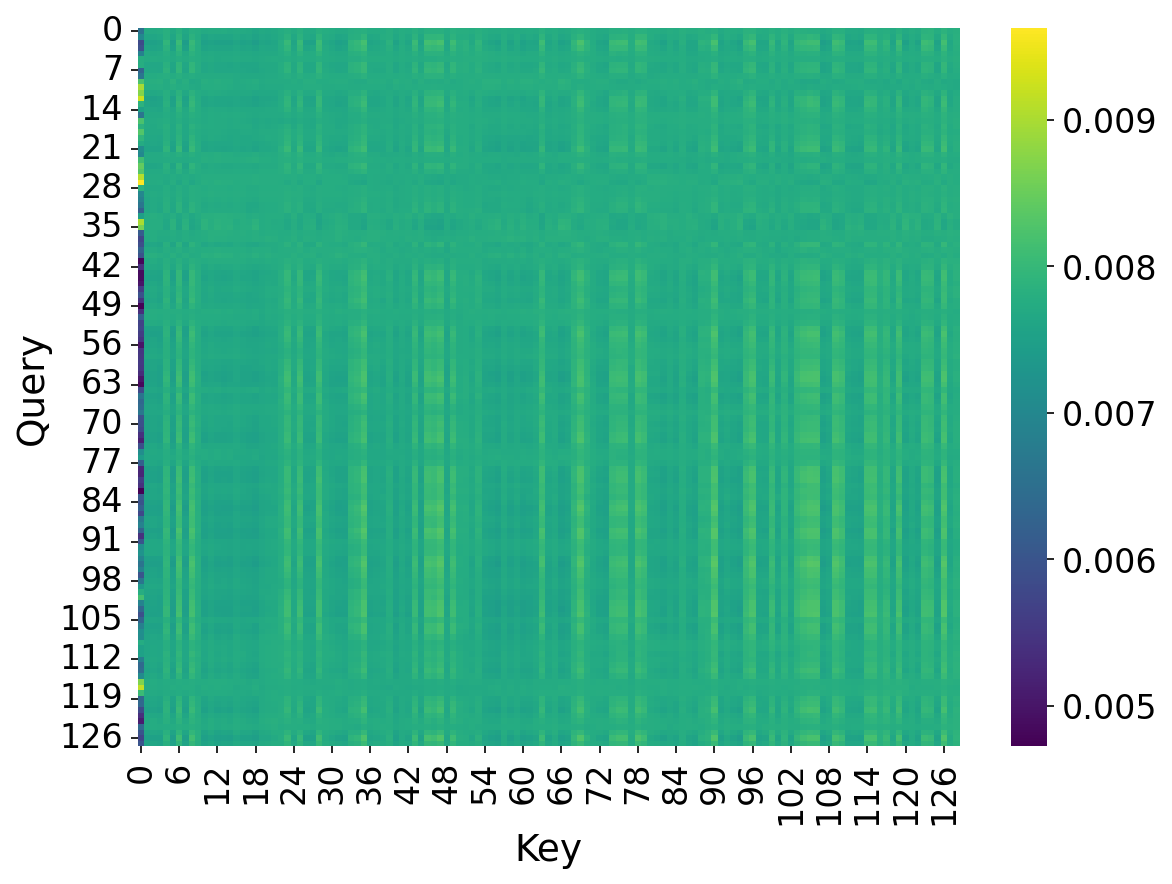}%
        \label{fig:att_as_bf}
    }
    \hfill
    \subfloat[\footnotesize \textnormal{Bona fide (Fusion)}]{%
        \includegraphics[width=0.3 \linewidth]{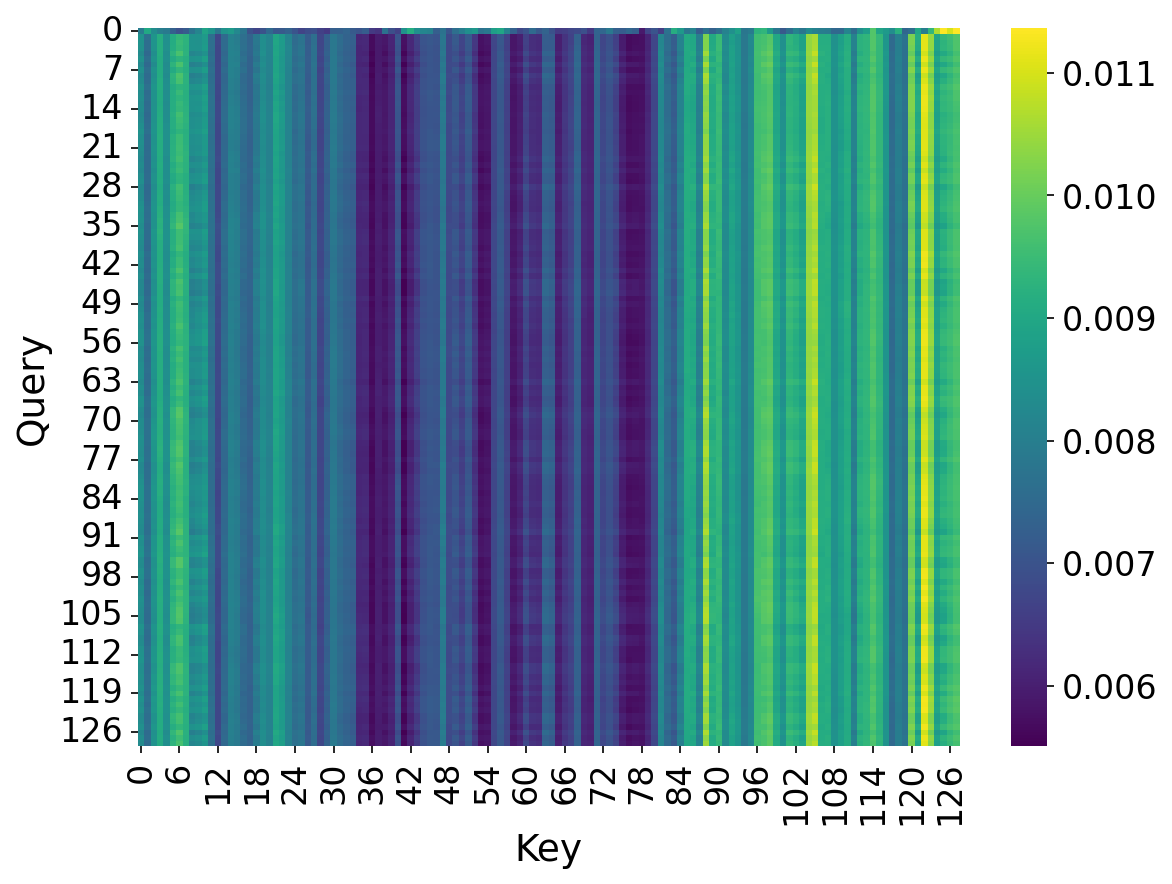}%
        \label{fig:att_fus_bf}
    }
    \vspace{-1em}
    \subfloat[\footnotesize \textnormal{Encodec (S$\rightarrow$A)}]{%
        \includegraphics[width=0.3 \linewidth]{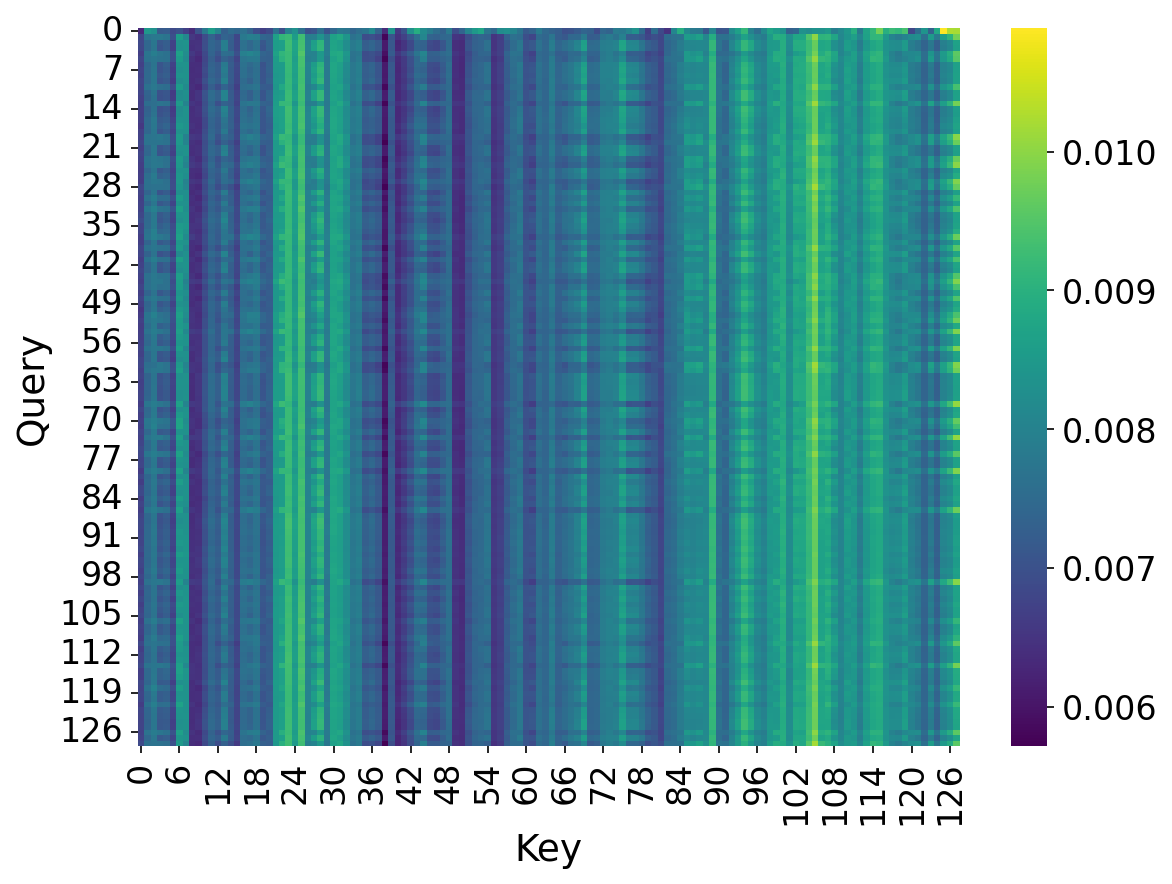}%
        \label{fig:att_sa_ec}
    }
    \hfill
    \subfloat[\footnotesize \textnormal{Encodec (A$\rightarrow$S)}]{%
        \includegraphics[width=0.3 \linewidth]{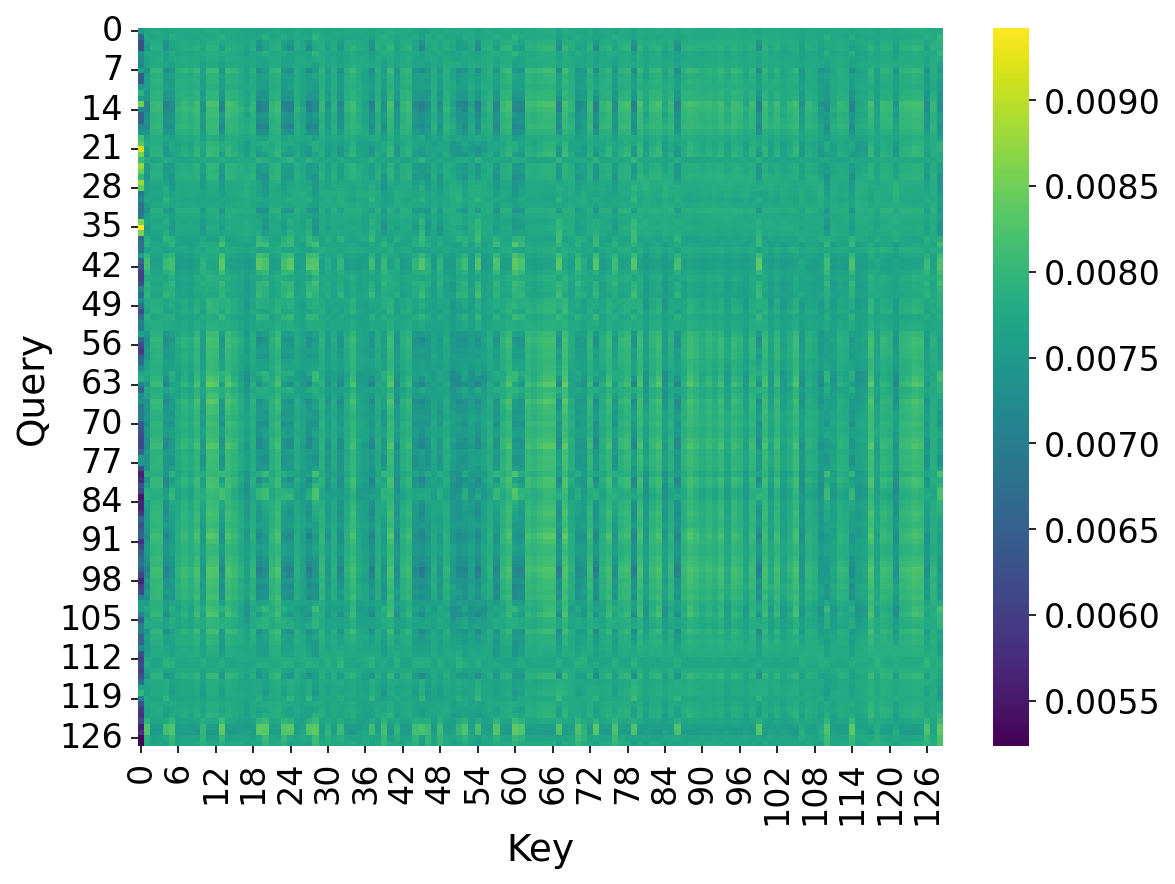}%
        \label{fig:att_as_ec}
    }
    \hfill
    \subfloat[\footnotesize \textnormal{Encodec (Fusion)}]{%
        \includegraphics[width=0.3 \linewidth]{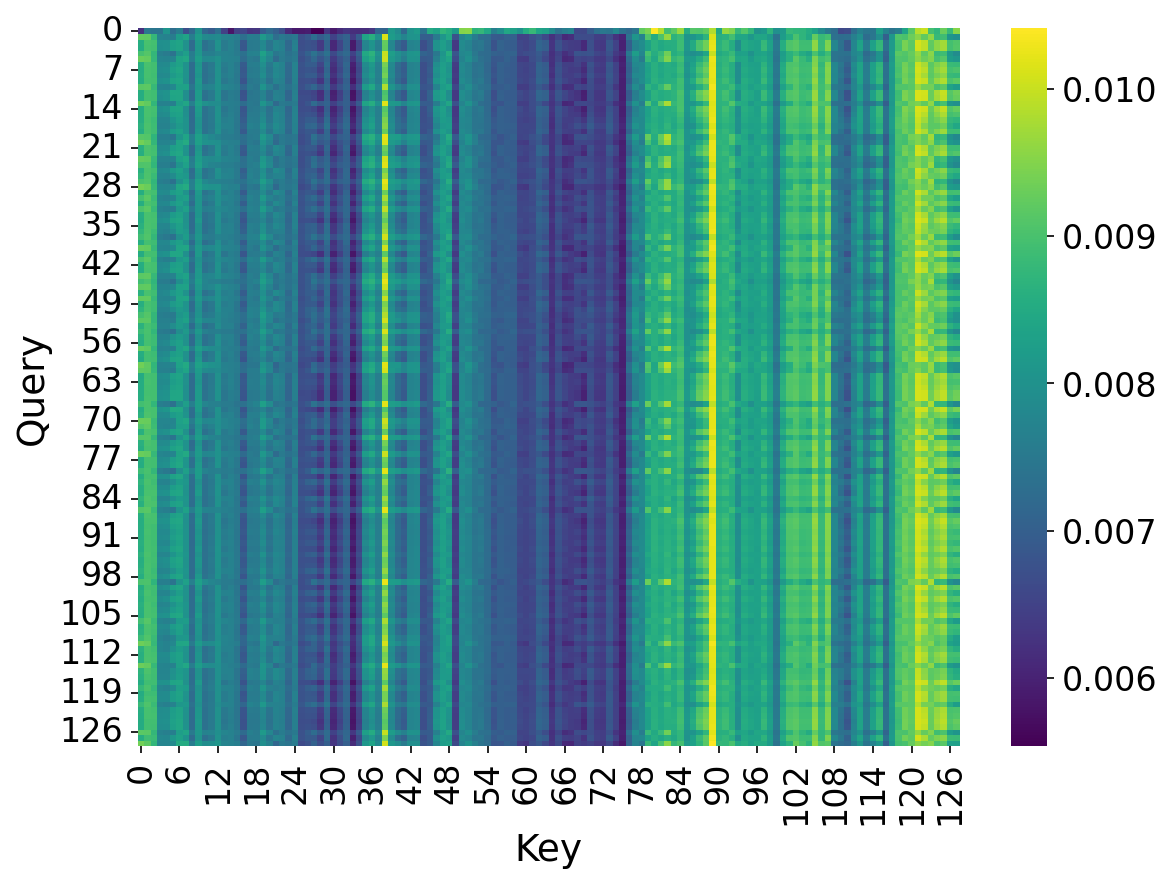}%
        \label{fig:att_fus_ec}
    }
    \caption{The Semantic-Acoustic Interaction Attention Maps for Case Study.}
    \vspace{-3mm}
    \label{fig:attention_map_comparison}
\end{figure}

%% file: sections/7.conclusion.tex
\section{Conclusion}
To sum up, we have shown that models trained on neural codec re-synthesised speech often overfit to silence regions and fail to generalize across diverse silence conditions or unseen content. 
To overcome these limitations, we introduced SASTNet, which combines a semantic encoder with a coarse‐to‐fine acoustic encoder to simultaneously preserve linguistic content and extract detailed codec fingerprints. Empirical results confirm that SASTNet substantially improves performance in codec‐based deepfake source tracing, demonstrating its effectiveness and robustness.

%% file: main.bbl
\begin{thebibliography}{10}
\providecommand{\url}[1]{#1}
\csname url@samestyle\endcsname
\providecommand{\newblock}{\relax}
\providecommand{\bibinfo}[2]{#2}
\providecommand{\BIBentrySTDinterwordspacing}{\spaceskip=0pt\relax}
\providecommand{\BIBentryALTinterwordstretchfactor}{4}
\providecommand{\BIBentryALTinterwordspacing}{\spaceskip=\fontdimen2\font plus
\BIBentryALTinterwordstretchfactor\fontdimen3\font minus \fontdimen4\font\relax}
\providecommand{\BIBforeignlanguage}[2]{{%
\expandafter\ifx\csname l@#1\endcsname\relax
\typeout{** WARNING: IEEEtran.bst: No hyphenation pattern has been}%
\typeout{** loaded for the language `#1'. Using the pattern for}%
\typeout{** the default language instead.}%
\else
\language=\csname l@#1\endcsname
\fi
#2}}
\providecommand{\BIBdecl}{\relax}
\BIBdecl

\bibitem{li2024audio}
M.~Li, Y.~Ahmadiadli, and X.-P. Zhang, ``A survey on speech deepfake detection,'' \emph{ACM Comput. Surv.}, vol.~57, no.~7, 2025.

\bibitem{wu2023defender}
H.~Wu, J.~Kang, L.~Meng, H.~Meng, and H.-y. Lee, ``The defender's perspective on automatic speaker verification: An overview,'' \emph{arXiv preprint arXiv:2305.12804}, 2023.

\bibitem{wu15e_interspeech}
Z.~Wu, T.~Kinnunen, N.~Evans, J.~Yamagishi, C.~Hanilçi, M.~Sahidullah, and A.~Sizov, ``Asvspoof 2015: the first automatic speaker verification spoofing and countermeasures challenge,'' in \emph{Proc. Interspeech}, 2015.

\bibitem{kinnunen17_interspeech}
T.~Kinnunen, M.~Sahidullah, H.~Delgado, M.~Todisco, N.~Evans, J.~Yamagishi, and K.~A. Lee, ``The asvspoof 2017 challenge: Assessing the limits of replay spoofing attack detection,'' in \emph{Proc. Interspeech}, 2017.

\bibitem{todisco2019asvspoof}
M.~Todisco, X.~Wang, V.~Vestman, M.~Sahidullah, H.~Delgado, A.~Nautsch, J.~Yamagishi, N.~Evans, T.~H. Kinnunen, and K.~A. Lee, ``{ASVspoof 2019: future horizons in spoofed and fake audio detection},'' in \emph{Proc. Interspeech}, 2019.

\bibitem{Liu_2023}
X.~Liu, X.~Wang, M.~Sahidullah, J.~Patino, H.~Delgado, T.~Kinnunen, M.~Todisco, J.~Yamagishi, N.~Evans, A.~Nautsch, and K.~A. Lee, ``Asvspoof 2021: Towards spoofed and deepfake speech detection in the wild,'' \emph{IEEE Transactions on Audio, Speech and Language Processing}, vol.~31, 2023.

\bibitem{Wang2024_ASVspoof5}
X.~Wang, H.~Delgado, H.~Tak, J.-w. Jung, H.-j. Shim, M.~Todisco \emph{et~al.}, ``{ASVspoof 5}: {Crowdsourced} speech data, deepfakes, and adversarial attacks at scale,'' in \emph{Proc. ASVspoof Workshop}, 2024.

\bibitem{yi2022add}
J.~Yi, R.~Fu, J.~Tao, S.~Nie, H.~Ma, C.~Wang, T.~Wang \emph{et~al.}, ``Add 2022: the first audio deep synthesis detection challenge,'' in \emph{Proc. ICASSP}, 2022.

\bibitem{yi2024add2023}
J.~Yi, C.~Y. Zhang, J.~Tao, C.~Wang, X.~Yan, Y.~Ren, H.~Gu, and J.~Zhou, ``Add 2023: Towards audio deepfake detection and analysis in the wild,'' \emph{arXiv preprint arXiv:2408.04967}, 2024.

\bibitem{borrelli2021synthetic}
C.~Borrelli, P.~Bestagini, F.~Antonacci, A.~Sarti, and S.~Tubaro, ``Synthetic speech detection through short-term and long-term prediction traces,'' \emph{EURASIP Journal on Information Security}, vol. 2021, 2021.

\bibitem{Yan2022AnII}
X.~Yan, J.~Yi, J.~Tao, C.~Wang, H.~Ma, T.~Wang, S.~Wang, and R.~Fu, ``An initial investigation for detecting vocoder fingerprints of fake audio,'' \emph{Proceedings of the 1st International Workshop on Deepfake Detection for Audio Multimedia}, 2022.

\bibitem{Zhang2023DistinguishingNS}
C.~Y. Zhang, J.~Yi, J.~Tao, C.~Wang, and X.~Yan, ``Distinguishing neural speech synthesis models through fingerprints in speech waveforms,'' in \emph{China National Conference on Chinese Computational Linguistics}, 2023.

\bibitem{zhu2022source}
T.~Zhu, X.~Wang, X.~Qin, and M.~Li, ``Source tracing: detecting voice spoofing,'' in \emph{Proc. APSIPA ASC}, 2022.

\bibitem{klein24_interspeech}
N.~Klein, T.~Chen, H.~Tak, R.~Casal, and E.~Khoury, ``Source tracing of audio deepfake systems,'' in \emph{Proc. Interspeech}, 2024.

\bibitem{arora2025landscape}
S.~Arora, K.-W. Chang, C.-M. Chien, Y.~Peng, H.~Wu, Y.~Adi, E.~Dupoux, H.-Y. Lee, K.~Livescu, and S.~Watanabe, ``On the landscape of spoken language models: A comprehensive survey,'' \emph{arXiv preprint arXiv:2504.08528}, 2025.

\bibitem{xie2025neural}
Y.~Xie, X.~Wang, Z.~Wang, R.~Fu, Z.~Wen, S.~Cao, L.~Ma, C.~Li, H.~Cheng, and L.~Ye, ``Neural codec source tracing: Toward comprehensive attribution in open-set condition,'' \emph{arXiv preprint arXiv:2501.06514}, 2025.

\bibitem{mishra2025towards}
J.~Mishra, M.~Chhibber, H.-j. Shim, and T.~H. Kinnunen, ``Towards explainable spoofed speech attribution and detection: a probabilistic approach for characterizing speech synthesizer components,'' \emph{arXiv preprint arXiv:2502.04049}, 2025.

\bibitem{negroni2025source}
V.~Negroni, D.~Salvi, P.~Bestagini, and S.~Tubaro, ``Source verification for speech deepfakes,'' \emph{arXiv preprint arXiv:2505.14188}, 2025.

\bibitem{xiao2025listen}
Y.~Xiao and R.~K. Das, ``Listen, analyze, and adapt to learn new attacks: An exemplar-free class incremental learning method for audio deepfake source tracing,'' \emph{arXiv preprint arXiv:2505.14601}, 2025.

\bibitem{phukan2025towards}
O.~C. Phukan, M.~M. Akhtar, A.~B. Buduru, R.~Sharma \emph{et~al.}, ``Towards neural audio codec source parsing,'' \emph{arXiv preprint arXiv:2506.12627}, 2025.

\bibitem{wu24p_interspeech}
H.~Wu, Y.~Tseng, and H.~yi~Lee, ``{CodecFake}: Enhancing anti-spoofing models against deepfake audios from codec-based speech synthesis systems,'' in \emph{Proc. Interspeech}, 2024.

\bibitem{chen2025codecfake+}
X.~Chen, J.~Du, H.~Wu, L.~Zhang, I.~Lin, I.~Chiu, W.~Ren, Y.~Tseng, Y.~Tsao, J.-S.~R. Jang, and H.-y. Lee, ``{CodecFake+}: A large-scale neural audio codec-based deepfake speech dataset,'' \emph{arXiv preprint arXiv:2501.08238}, 2025.

\bibitem{chen2025codec}
X.~Chen, I.~Lin, L.~Zhang, J.~Du, H.~Wu, H.-y. Lee, J.-S.~R. Jang \emph{et~al.}, ``Codec-based deepfake source tracing via neural audio codec taxonomy,'' \emph{arXiv preprint arXiv:2505.12994}, 2025.

\bibitem{guo2025recent}
Y.~Guo, Z.~Li, H.~Wang, B.~Li, C.~Shao, H.~Zhang, C.~Du, X.~Chen, S.~Liu, and K.~Yu, ``Recent advances in discrete speech tokens: A review,'' \emph{arXiv preprint arXiv:2502.06490}, 2025.

\bibitem{wu2024towards}
H.~Wu, X.~Chen, Y.-C. Lin, K.-w. Chang, H.-L. Chung, A.~H. Liu, and H.-y. Lee, ``Towards audio language modeling-an overview,'' \emph{arXiv preprint arXiv:2402.13236}, 2024.

\bibitem{wu2024codec}
H.~Wu, H.-L. Chung, Y.-C. Lin, Y.-K. Wu, X.~Chen, Y.-C. Pai \emph{et~al.}, ``Codec-{SUPERB}: An in-depth analysis of sound codec models,'' in \emph{Findings Assoc. Comput. Linguist.}, 2024.

\bibitem{wu2024codec_slt24}
H.~Wu, X.~Chen, Y.-C. Lin, K.~Chang, J.~Du, K.-H. Lu \emph{et~al.}, ``{Codec-SUPERB@ SLT 2024}: A lightweight benchmark for neural audio codec models,'' in \emph{Proc. IEEE Spoken Lang. Technol. Workshop}, 2024.

\bibitem{kawa23b_interspeech}
P.~Kawa, M.~Plata, M.~Czuba, P.~Szymański, and P.~Syga, ``Improved deepfake detection using whisper features,'' in \emph{Interspeech 2023}, 2023.

\bibitem{jung2022aasist}
J.-w. Jung, H.-S. Heo, H.~Tak, H.-j. Shim, J.~S. Chung, B.-J. Lee, H.-J. Yu, and N.~Evans, ``Aasist: Audio anti-spoofing using integrated spectro-temporal graph attention networks,'' in \emph{ICASSP 2022-2022 IEEE international conference on acoustics, speech and signal processing (ICASSP)}, 2022.

\bibitem{radford2022whisper}
A.~Radford, J.~W. Kim, T.~Xu, G.~Brockman, C.~McLeavey, and I.~Sutskever, ``Robust speech recognition via large-scale weak supervision,'' 2022.

\bibitem{babu2021xls}
A.~Babu, C.~Wang, A.~Tjandra \emph{et~al.}, ``Xls-r: Self-supervised cross-lingual speech representation learning at scale,'' \emph{arXiv preprint arXiv:2111.09296}, 2021.

\bibitem{huang2022masked}
P.-Y. Huang, H.~Xu, J.~Li, A.~Baevski, M.~Auli, W.~Galuba, F.~Metze, and C.~Feichtenhofer, ``Masked autoencoders that listen,'' \emph{Advances in Neural Information Processing Systems}, vol.~35, pp. 28\,708--28\,720, 2022.

\bibitem{he2021mae}
K.~He, X.~Chen, S.~Xie, Y.~Li, P.~Doll{\'a}r, and R.~Girshick, ``Mae: Masked autoencoders are scalable vision learners,'' \emph{arXiv preprint arXiv:2111.06377}, 2021.

\bibitem{wu2022adversarial}
H.~Wu, P.-C. Hsu, J.~Gao, S.~Zhang, S.~Huang, J.~Kang, Z.~Wu, H.~Meng, and H.-Y. Lee, ``Adversarial sample detection for speaker verification by neural vocoders,'' in \emph{ICASSP 2022 - 2022 IEEE International Conference on Acoustics, Speech and Signal Processing (ICASSP)}, 2022.

\bibitem{chen24p_interspeech}
X.~Chen, J.~Du, H.~Wu, J.-S.~R. Jang, and H.~yi~Lee, ``Neural codec-based adversarial sample detection for speaker verification,'' in \emph{Interspeech 2024}, 2024.

\bibitem{du2024towards}
R.~Du, J.~Yao, Q.~Kong, and Y.~Cao, ``Towards out-of-distribution detection in vocoder recognition via latent feature reconstruction,'' \emph{arXiv preprint arXiv:2406.02233}, 2024.

\bibitem{chen2022push}
X.~Chen, H.~Wu, H.~Meng, H.-y. Lee, and J.-S.~R. Jang, ``Push-pull: Characterizing the adversarial robustness for audio-visual active speaker detection,'' in \emph{2022 IEEE Spoken Language Technology Workshop (SLT)}, 2023.

\bibitem{chen2024mtd}
X.~Chen, H.~Wu, C.-C. Wang, H.-Y. Lee, and J.-S.~R. Jang, ``Multimodal transformer distillation for audio-visual synchronization,'' in \emph{ICASSP 2024 - 2024 IEEE International Conference on Acoustics, Speech and Signal Processing (ICASSP)}, 2024.

\bibitem{chen2025localizingaudiovisualdeepfakeshierarchical}
X.~Chen, S.-P. Cheng, J.~Du \emph{et~al.}, ``Localizing audio-visual deepfakes via hierarchical boundary modeling,'' \emph{arXiv preprint arXiv:2508.02000}, 2025.

\bibitem{vaswani2017attention}
A.~Vaswani, N.~Shazeer, N.~Parmar, J.~Uszkoreit, L.~Jones, A.~N. Gomez, {\L}.~Kaiser, and I.~Polosukhin, ``Attention is all you need,'' \emph{Advances in neural information processing systems}, vol.~30, 2017.

\bibitem{liebel2018auxiliary}
L.~Liebel and M.~K{\"o}rner, ``Auxiliary tasks in multi-task learning,'' \emph{arXiv preprint arXiv:1805.06334}, 2018.

\bibitem{tak2022automatic}
H.~Tak, M.~Todisco, X.~Wang, J.-w. Jung, J.~Yamagishi, and N.~Evans, ``Automatic speaker verification spoofing and deepfake detection using wav2vec 2.0 and data augmentation,'' in \emph{Proc. Odyssey Speaker Lang. Recognit. Workshop}, 2022.

\bibitem{tak2022rawboost}
H.~Tak, M.~Kamble, J.~Patino, M.~Todisco, and N.~Evans, ``Rawboost: A raw data boosting and augmentation method applied to automatic speaker verification anti-spoofing,'' in \emph{Proc. ICASSP}, 2022.

\bibitem{yamagishi2019cstr}
J.~Yamagishi, C.~Veaux, K.~MacDonald \emph{et~al.}, ``Cstr vctk corpus: English multi-speaker corpus for cstr voice cloning toolkit (version 0.92),'' \emph{Univ. of Edinburgh, The Centre for Speech Technology Research (CSTR)}, 2019.

\bibitem{pmlr-v235-yang24x}
D.~Yang, J.~Tian, X.~Tan, R.~Huang, S.~Liu, H.~Guo, X.~Chang, J.~Shi, S.~Zhao, J.~Bian, Z.~Zhao, X.~Wu, and H.~M. Meng, ``{U}ni{A}udio: Towards universal audio generation with large language models,'' in \emph{Proceedings of the 41st International Conference on Machine Learning}, 2024.

\bibitem{wang2024maskgct}
Y.~Wang, H.~Zhan, L.~Liu, R.~Zeng, H.~Guo, J.~Zheng \emph{et~al.}, ``Mask{GCT}: Zero-shot text-to-speech with masked generative codec transformer,'' in \emph{Proc. ICLR}, 2025.

\bibitem{zhang2023impact}
Y.~Zhang, Z.~Li, J.~Lu, H.~Hua, W.~Wang, and P.~Zhang, ``The impact of silence on speech anti-spoofing,'' \emph{IEEE/ACM Transactions on Audio, Speech, and Language Processing}, vol.~31, pp. 3374--3389, 2023.

\bibitem{zhang122021effect}
Y.~Zhang, W.~Wang, and P.~Zhang, ``The effect of silence and dual-band fusion in anti-spoofing system,'' in \emph{Proc. Interspeech}, 2021, pp. 4279--4283.

\bibitem{wang2024genuine}
X.~Wang, R.~Fu, Z.~Wen, Z.~Wang, Y.~Xie, Y.~Liu, J.~Tao, X.~Liu, Y.~Li, X.~Qi \emph{et~al.}, ``Genuine-focused learning using mask autoencoder for generalized fake audio detection,'' \emph{arXiv preprint arXiv:2406.03247}, 2024.

\bibitem{huang2024dynamic}
C.-y. Huang, K.-H. Lu, S.-H. Wang, C.-Y. Hsiao, C.-Y. Kuan, H.~Wu, S.~Arora, K.-W. Chang, J.~Shi, Y.~Peng \emph{et~al.}, ``Dynamic-superb: Towards a dynamic, collaborative, and comprehensive instruction-tuning benchmark for speech,'' in \emph{ICASSP 2024-2024 IEEE International Conference on Acoustics, Speech and Signal Processing (ICASSP)}.\hskip 1em plus 0.5em minus 0.4em\relax IEEE, 2024, pp. 12\,136--12\,140.

\bibitem{huang2025dynamicsuperb}
C.-Y. Huang \emph{et~al.}, ``Dynamic-{SUPERB} phase-2: A collaboratively expanding benchmark for measuring the capabilities of spoken language models with 180 tasks,'' in \emph{Proc. ICLR}, 2025.

\bibitem{yang2025towards}
C.-K. Yang, N.~S. Ho, and H.-y. Lee, ``Towards holistic evaluation of large audio-language models: A comprehensive survey,'' \emph{arXiv preprint arXiv:2505.15957}, 2025.

\bibitem{lin2025preliminary}
Y.-X. Lin, C.-K. Yang, W.-C. Chen \emph{et~al.}, ``A preliminary exploration with gpt-4o voice mode,'' \emph{arXiv preprint arXiv:2502.09940}, 2025.

\end{thebibliography}
